\documentclass[12pt, a4paper]{article}

\usepackage[OT1]{fontenc}
\usepackage{amsthm,amsmath}

\usepackage[utf8]{inputenc} 

\usepackage{dsfont} 
\usepackage{array}
\usepackage{graphicx}
\usepackage{epstopdf}
\usepackage[dvipsnames]{xcolor}
\usepackage{subfigure} 

\usepackage{multirow}
\usepackage[backend=bibtex,
		bibencoding=utf8,
	    sorting=nyt,
	    style=numeric
	    ]{biblatex}

\usepackage{amssymb} 
\usepackage{amsfonts}
\usepackage{enumitem}

\usepackage{textcomp} 
\usepackage{calc}

\usepackage{bm}

\usepackage{hyperref}
\hypersetup{colorlinks=true,linkcolor=blue,urlcolor=blue, citecolor=blue}

\numberwithin{equation}{section}
\theoremstyle{plain}
\newtheorem{thm}{Theorem}[section]

\newcommand{\RR}{\mathbb{R}} 
\newcommand{\NN}{\mathbb{N}} 
\newcommand{\ZZ}{\mathbb{Z}} 
\newcommand{\EE}{\mathbb{E}} 
\newcommand{\VV}{\mathrm{Var}} 
\newcommand{\Cov}{\mathrm{Cov}} 

\newcommand{\KK}{\mathcal{K}}

\newcommand{\XX}{\mathbf{X}} 
\newcommand{\YY}{\mathbf{Y}} 

\newcommand{\de}{\mathrm{d}} 
\newcommand{\e}[1]{\mathrm{e}^{#1}} 
\newcommand{\ii}{\mathrm{i}} 
\newcommand{\ind}[1]{\mathds{1}\left[#1\right]} 

\newcommand{\abs}[1]{\left|#1\right|}
\newcommand{\norm}[1]{\left\|#1\right\|}
\newcommand{\floor}[1]{\left\lfloor#1\right\rfloor}

\newcommand{\I}[2]{\left[#1{,}#2\right]}

\newcommand{\DPP}{\mathrm{DPP}}
\newtheorem{proposition}{Proposition}[section]
\newtheorem{corollary}{Corollary}[section]
\newtheorem{lemma}{Lemma}[section]

\newcolumntype{M}[1]{>{\centering} m{#1}} 

\graphicspath{{./Res_Simu/}}
\begin{document}

\title{Monte Carlo integration of non-differentiable functions on $[0{,}1]^\iota$, $\iota=1,\dots,d$ , using a single determinantal point pattern defined on $[0{,}1]^d$}
\author{Jean-François Coeurjolly, Adrien Mazoyer and Pierre-Olivier Amblard}

\maketitle

\begin{abstract}
This paper concerns the use of a particular class of determinantal point processes (DPP), a class of repulsive spatial point processes, for Monte Carlo integration.
Let $d\ge 1$, $I\subseteq \overline d=\{1,\dots,d\}$ with $\iota=|I|$. Using a single set of $N$  quadrature points $\{u_1,\dots,u_N\}$ defined, once for all, in dimension $d$ from the realization of {a specific} DPP, we investigate ``minimal'' assumptions on the integrand in order to obtain unbiased Monte Carlo estimates  of  $\mu(f_I)=\int_{[0,1]^\iota} f_I(u) \de u$  for any known $\iota$-dimensional integrable function on $[0,1]^\iota$. In particular, we show that the resulting estimator has variance with order $N^{-1-(2s\wedge 1)/d}$ when the integrand belongs to some Sobolev space with regularity $s > 0$. When $s>1/2$ (which includes a large class of non-differentiable functions), the variance is asymptotically explicit and the estimator is shown to satisfy a Central Limit Theorem.
\end{abstract}

\section*{Introduction}

{The paper investigates Monte-Carlo evaluation of the integral $\mu(f)=\int_{[0,1]^\iota} f(u) \de u$    for a {\em  known} $\iota$-dimensional integrable function on $[0,1]^\iota$ using {\em a single set} of $N$  quadrature points $\{u_1,\dots,u_N\}$ defined once for all in dimension $d\geq \iota$.
The same set of nodes, defined in $[0,1]^d$, may therefore be used to estimate {\em a finite number of different integrals}, and therefore this set does not exploit the form of $f$, (locations of its possible singularities, {\it etc})}

Such an approach may be of importance in any application where repeatedly integrating a function over different subsets is needed, ({\it e.g. } in sensitivity analysis using experimental designs) or where calculating marginals is required ({\it e.g. }  for evaluating the marginal likelihood in some parametric statistical models).  The first example is emblematic of our motivation. 

In the context of computer experiments (see for example \cite[Chapter 5]{Santneretal13}), complex phenomena are simulated using a mathematical model to replace the process which generates the data. Usually, the model depends on a large number of parameters (inputs). An objective of the experiments is to quantify the influence of the variability of the inputs on the variable of interest. An experiment consists in running simulations, where each simulation represents a possible combination of the inputs. It is impossible in practice to consider all possible configurations, the number of simulations being limited. Therefore, the design of experiments, {\it i.e. } the choice of  combinations of  inputs, is of great importance. Under a lack of information on how inputs are linked to outputs, a strategy is to spread chosen inputs to cover as much as possible all the input space. This technique is called \textit{space-filling design}. It can be summarized by generating~$N$ points in a given space which regularly cover this space. 
Latin hypercubes \cite{McKayetal79,Owen92}, low discrepancy sequences (see {\it e.g. }\cite{Halton64, Sobol67}) are standard methods to generate designs. The goal of computer experiments is not only to examine the influence of all the inputs on an output of interest, but also the influence of a subset of these inputs, or also the influence of a particular combination of subsets of these inputs. Since computer experiments may be very expensive in terms of computation load and/or storage capacity, the regularity of the coverage of the designs should be conserved when the initial configuration is projected onto lower dimensional spaces. This would allow to use the initial configuration to study the influence of subsets of  inputs for example with the same efficiency. { Furthermore, integrating a function over different subsets of its variables may be of importance in this context, for example in sensitivity analysis  \cite[Chapter 7]{Santneretal13}. When integrating explicitly is impossible, numerical methods are used. We turn our attention here to Monte Carlo integration.}


Monte Carlo integration has a long history and it is not the aim of this paper to make a detailed bibliography. We refer the interested reader by an extensive treatment and bibliography to the electronic book by \cite{owen:book:13}.
Let us however cite a few methods keeping in mind what we mean by ``minimal'' assumptions in the situation $\iota=d$. {Crude} Monte Carlo methods and importance sampling methods (see {\it e.g. }\cite{RobertCasella04}) consist in using i.i.d. nodes $\{u_1,\dots,u_N\}$ with a so-called proposal density (the uniform density in the usual situation {in which the target distribution is also uniform}). Under some $L^2([0,1]^d)$ type assumption on the integrand, the resulting estimator denoted by 
{$\widehat \mu_N(f)$}
has a variance proportional to $N^{-1}$ and satisfies a Central Limit Theorem. When $d$ is large, Monte Carlo Markov Chains (MCMC) methods, where the set of quadrature points is the realization of a particular Markov chain, are usually preferred. When $f\in L^1([0,1]^d)$, the variance of 
{$\widehat \mu_N(f)$}
is still of order $N^{-1}$ and satisfies a CLT (see {\it e.g. }\cite{douc:moulines:stoffer:14}). To improve the rate of convergence, the price to pay is to require some regularity assumptions on 
{$f$.}
Many methods exist in the literature: grid-based stratified methods \cite{haber:66}, possibly combined with antithetic sampling (see~\cite[Chapter~10]{owen:book:13}), Quasi Monte Carlo and randomized versions, scrambled nets \cite{dick:etal:13,owen:97,owen:08,owen:book:13,basu:mukherjee:17}, {\it etc}. For example, a version of scrambled nets with antithetic sampling can lead to an estimator with variance $\mathcal O(N^{-3-2/d} \log(N)^{d-1})$ if, to simplify Owen's assumption \cite{owen:08}, {$f$} is $d$ times continuously differentiable on $[0,1]^d$. Additional assumptions on the scrambled net are required to obtain a CLT.
Grid-based stratified methods which are maybe the first simple alternative to ordinary Monte Carlo methods require that {$f$} is continuously differentiable on $[0,1]^d$ and yield an estimator satisfying a CLT with variance  asymptotically proportional to $N^{-1-2/d}$. Let us also mention that \cite{owen:97} showed that a version of scrambled net has a variance $o(N^{-1})$ under the sole assumption that 
{$f\in L^2([0,1]^d)$}
 but the rate is not explicit until strong regularity assumptions are made on 
 {$f$.}

In a recent work, \cite{BardenetHardy20} proposed another alternative by defining the nodes as the realization of a repulsive point pattern, and in particular a determinantal point pattern. The class of determinantal point processes (DPPs for short) has received a growing attention in the last decades (see {\it e.g. }\cite{Soshnikov00, ShiraiTakahashi03, Houghetal09, Lavancieretal15_1, Decreusefondetal16}), thanks to its very appealing properties in particular in terms of tractability and exact simulation. \cite{BardenetHardy20} have defined an Orthogonal Polynomial Ensemble, which is a particular inhomogeneous DPP whose kernel is defined through orthonormal polynomials. Under the assumption that 
{$f$}
 is continuously differentiable and compactly supported in $B^\prime \subset [0,1]^d$ the authors obtained an estimator with variance equivalent to an explicit constant times $N^{-1-1/d}$. 

In this paper, we investigate a different DPP. To be more explicit, we consider the most natural kernel, called the Dirichlet kernel in this paper, which is based on the Fourier decomposition of a rectangular subset of $N$ indices of $\ZZ^d$. It is a projection DPP {that is a point process} which produces almost surely $N$ points. It has the advantage to lead to a homogeneous DPP pattern, an interesting characteristic as we want the pattern to be used to estimate any integral {without taking advantage of the integrand $f$}. A second advantage is that the marginals are fully characterized and explicit which means that marginals can efficiently be used to estimate $\widehat \mu_N(f_I)$ for any $I \subset \overline d=\{1,\dots,d\}$. Last but not least, our main result Theorem~\ref{thm:asymp_Iest}, shows that the resulting estimator $\widehat \mu_N(f_I)$ has asymptotic variance proportional to $N^{-1-(2s \wedge 1)/d}$ for any $f_I \in \mathcal H^s([0,1]^\iota)$ where $\mathcal H^s([0,1]^\iota)$ is some Sobolev space with regularity $s\ge 0$, see~\eqref{eq:defHs} for more details. We remind that for periodic functions $L^2([0,1]^\iota)= \mathcal H^0([0,1]^\iota)$ and if $f_I$ is periodic and continuously differentiable then $f_I\in \mathcal H^1([0,1]^\iota)$. In particular, our result states that when $s>1/2$ (thus potentially for non-differentiable functions), the variance is asymptotically equivalent to an explicit constant times $N^{-1-1/d}$. In this case, we also obtain a central limit theorem for the estimator (assuming in addition that the integrand is bounded). As a summary, the estimator proposed has the characteristic to exhibit a variance that decreases faster than the ordinary Monte Carlo as soon as $s>0$. The decay is slower than methods such as grid-based methods, scrambled nets, {\it etc}, but require much less regularity assumptions and can be applied to any $\iota$-dimensional function, $\iota=1,\dots,d$.

The  paper is organized as follows. Section \ref{sec:backgr} contains a background on spatial point processes, generalities on the projection of spatial point processes and DPPs. We also outline the interest of repulsive point processes and in particular DPPs for Monte Carlo integration. Section~\ref{sec:dirichletDPP} introduces the Dirichlet DPP and exposes some of its properties. Our main result, Theorem~\ref{thm:asymp_Iest}, is presented in Section~\ref{sec:MC}. It details convergence results for Monte Carlo integration based on the realization of a Dirichlet DPP. A multivariate version of this CLT is also proposed. Section~\ref{sec:exp} discusses computational aspects for the simulation of Dirichlet DPPs and contains a simulation study which illustrates our results. 
{In Section \ref{sec:comparisons}, we perform a deeper numerical comparison of the sampling used here with several existing  methods among which Bardenet\&Hardy's, stratified sampling, maximinLHS, {\it etc}}.  
Finally, all proofs of the results are postponed to Appendix~\ref{sec:appendix}.


\section{Background and notation} \label{sec:backgr}

\subsection{Spatial point processes}

A spatial point process~$\XX$ defined on a Borel set~$B\subseteq\RR^d$ is a locally finite measure on $B$, see for example \cite{MollerWaagepeterson04} and references therein for measure theoretical details, whose realization is of the form 
$\{x^{(1)}, \ldots , x^{(k)}\}\in B^k$
where~$k$ is the realization of a random variable and the~$x^{(i)}$'s represent the events. We assume that $\XX$ is simple meaning that two events cannot occur at the same location. Thus, $\XX$ is viewed as a locally finite random set.

In most cases, the distribution of a point process $\XX$ can be described by its intensity functions {$\rho_\XX^{(k)}:B^k\rightarrow \RR^+$, $k\in\mathbb{N} \setminus \{0\}$. By Campbell theorem, see {\it e.g. }\cite{MollerWaagepeterson04}, $\rho_\XX^{(k)}$ is characterized by the following integral representation: for any non-negative measurable function $h:B^k \to \RR^+$
\begin{align}\label{eq:rhok_def}		
\EE\Bigg[&\sum_{x^{(1)}, \ldots , x^{(k)} \in \XX}^{\neq} h\left(x^{(1)}, \ldots , x^{(k)}\right)\Bigg]\\
&	 = \int_{B^k} \rho_\XX^{(k)}\left(x^{(1)},\ldots,x^{(k)}\right) h\left(x^{(1)}, \ldots , x^{(k)}\right) \de x^{(1)}\ldots\de x^{(k)}\nonumber
\end{align}  
where~$\neq$ over the summation means that~$x^{(1)}, \ldots , x^{(k)}~$ are pairwise distinct points. Intuitively, for any pairwise distinct points~\mbox{$x^{(1)},\ldots,x^{(k)}\in B$,} \linebreak$\rho_\XX^{(k)}\left(x^{(1)},\ldots,x^{(k)}\right)\de x^{(1)}\ldots\de x^{(k)}$ is the probability that $\XX$ has a point in each~of the $k$ infinitesimally small sets around~$x^{(1)},\ldots,x^{(k)}$ with volumes $\de x^{(1)},\ldots,\de x^{(k)}$, respectively. When $k=1$, this yields the intensity function  simply denoted  by $\rho_{\XX}=\rho_{\XX}^{(1)}$.
The second order intensity~$\rho_\XX^{(2)}$ is used to define the pair correlation function
	\begin{equation} \label{eq:pcf_def}
		g_\XX(x^{(1)},x^{(2)}) = \frac{\rho_\XX^{(2)}(x^{(1)},x^{(2)})}{\rho_\XX(x^{(1)})\rho_\XX(x^{(2)})}\,
 	\end{equation}
	for pairwise distinct $x^{(1)}$, $x^{(2)} \in B$ and
	where $g_\XX(x^{(1)},x^{(2)})$ is set to 0 if~$\rho_\XX(x^{(1)})$ or~$\rho_\XX(x^{(2)})$ is zero.
	 By convention, $\rho_\XX^{(k)}\left(x^{(1)},\ldots,x^{(k)}\right)$ is set to 0 if~$x^{(i)}=x^{(j)}$ for some~$i\neq j$. 
	Therefore~$g_\XX(x,x)$ is also set to 0 for all~$x\in B$ by convention. The pair correlation function (pcf for short) can be used to determine the local interaction between points of~$\XX$~located at $x$ and $y$:~$g_\XX(x,y) > 1$ characterizes positive correlation between the points;~$g_\XX(x,y)=1$ means there is no interaction (typically a Poisson point process);~$g_\XX(x,y) < 1$ characterizes negative correlations. A point pattern is often referred to as a repulsive point process, if $g_\XX(x,y)<1$ for any $x,y\in B$ (see e.g.~\cite[Section 6.5]{Illianetal08}).
Finally, a point process $\XX$ with constant intensity function on $B$ is said to be homogeneous. 

\subsection{Projection of a spatial point process} \label{ssec:projectionSPP}

In this work, we sometimes consider projection of spatial point processes. By projection, we mean that we keep a given number of coordinates from the original spatial point process. Such a framework requires that the original point process $\XX$ must be defined on a compact set $B \subset \RR^d$: otherwise, the configuration of  points of the projected point processes may not form a locally finite configuration, as also noticed in the two-dimensional case in \cite[p. 17]{Baddeley07}. 

This section presents a few notation in this context. Let $I\subseteq \overline{d} := \{1,\ldots, d\}$ with cardinality $|I|=\iota$. Let $B_1,\dots,B_d$ compact sets of $\RR$ and $B=B_1\times \dots\times B_d$, denote by~$B_I$ its orthogonal projection onto~$\RR^{\iota}$. In particular
	$
	B_I = \prod_{i\in I} B_i\,
	$
	with $B=B_{\overline{d}}$.
We denote by~$P_I$ the orthogonal projection of~$\RR^d$ onto~$\RR^{\iota}$. To ease the reading, we let $B_\ell=B_0$ for $\ell=1,\dots,d$ and even to fix ideas let $B_0=\I{0}{1}$. Thus, $B_I=B_0^\iota=\I{0}{1}^\iota$. For any $x \in B$, we let $x_I=P_I x$ and for a point process $\XX$ defined on $B$, the projected point process~$\XX_I=P_I\XX$ is then defined on~$B_I$. Intensity functions and Laplace functionals for $P_I \XX$ can be derived from the corresponding functions and functionals from $\XX$, see \cite{Mazoyeretal20} for more details and Section~\ref{ssec:ProjDirichlet} for the particular {point process} considered in this paper.


\subsection{Determinantal point processes} \label{ssec:DPP}
%

In this section, the class of continuous DPPs is introduced. Again, we restrict our attention to DPPs defined on a compact set $B\subset \RR^d$. A point process~$\XX$ on~$B$ is said to be a DPP on $B$ with kernel $K:B\times B \to \mathbb C$ if for any $k\ge 1$ its~$k$th order intensity function is given by
\begin{equation} \label{eq:rhok_DPP}
\rho_\XX^{(k)}\left(x^{(1)},\ldots,x^{(k)}\right) = \det\left[K\left(x^{(i)},x^{(j)}\right)\right]_{i,j=1}^k
\end{equation}
and we simply denote by $\XX\sim \DPP_B(K)$. Note that $K$ needs to be non-negative definite to ensure $\rho_\XX^{(k)}\geqslant0$. Our results rely on the spectral decomposition of~$K$, see~\eqref{eq:Mercer}. Therefore, we assume that $K$ is a continuous covariance function. The intensity of $\XX$ is given by $\rho_\XX(x)=K(x,x)$ and its pcf by
\begin{equation} \label{eq:pcf_DPP}
g_\XX(x,y) = 1-\frac{\left|K(x,y)\right|^2}{K(x,x)K(y,y)}.
\end{equation}
The popularity of DPPs relies mainly upon \eqref{eq:rhok_DPP}-\eqref{eq:pcf_DPP}: all moments of $\XX$ are explicit and since
$K$ is Hermitian, 
$g_\XX(x,y)<1$ for any $x,y\in B$.
The kernel $K$ defines an integral operator $\KK$ (see {\it e.g. }\cite{DebnathMikusinski05}) defined for any $f\in L^2(B)$ by
\[
\KK(f)(x) = \int_B K(x,y) f(y) \de y,\quad x\in B.
\]
From Mercer's Theorem \cite[Sec. 98]{RieszSznagy90}, $K$ admits the following spectral decomposition for any $x,y\in B$
\begin{equation} \label{eq:Mercer}
K(x,y) = \sum_{j\in \NN} \lambda_j \phi_j(x)\overline{\phi_j(y)}
\end{equation}
where
	$\{\phi_j\}_{j\in \NN}$ are eigenfunctions associated to $\KK$ and form an orthonormal basis of~$L^2(B)$,
and where
	$\{\lambda_j\}_{j\in \NN}$ are the eigenvalues of $\KK$ satisfying $\lambda_j\ge 0$ for any $j\in \NN$. We abuse notation in the sequel and refer $\lambda_j$'s to as the eigenvalues of $K$.

The existence of a DPP on $B$ with kernel $K$ is ensured if its eigenvalues satisfy $\lambda_j\leqslant1$ for any $j\in\NN$, see {\it e.g. }\cite[Theorem 4.5.5.]{Houghetal09}.
Eigenvalues and eigenfunctions are indexed here by $\NN$ in \eqref{eq:Mercer}, but other countable sets could be considered. In particular, the $d$-dimensional Fourier basis is indexed by $\ZZ^d$. 
A DPP $\XX\sim\DPP_B(K)$ is said to be homogeneous if $K$ is the restriction on $B\times B$ of a kernel $C$ defined on $\RR^d\times \RR^d$ which satisfies $C(x,y)=C({o},x-y)$ for any $x,y\in\RR^d$ where $o$ is the origin in $\RR^d$. In that case, we will refer to $K$ as a stationary kernel and will use the abusive notation $K(x,y)\equiv K(x-y)$. 

A kernel $K$ such that $\lambda_j \in \{0,1\}$ for $j\in \NN$ is called a ``projection kernel'' and the corresponding DPP a ``projection DPP''. The number of points in $B$ of such a {point process} is almost surely constant and equal to the number of non-zero eigenvalues of $K$ (see e.g.~\cite{Lavancieretal15_1}).

\subsection{Why  are DPPs interesting for Monte Carlo integration?}

The repulsive nature of DPPs can be exploited to generate quadrature points that explore nicely the input space.
To see this, let $B \subset \RR^d$ be a bounded set, $f\in L^2(B)$ and $\YY$  a homogeneous point process on $B$ with intensity parameter $\rho_\YY$ and pair correlation function $g_\YY$ (a similar result would hold in the inhomogeneous case). Campbell's Theorem \eqref{eq:rhok_def} ensures that the estimator
\begin{equation} \label{eq:est_I_DPP}
	\widehat{\mu}(f) = \frac1{\rho_\YY}  \sum_{u\in\YY} f(u)
\end{equation}
is an unbiased estimator of $\mu(f) = \int_B f(u) \de u$ with variance
\begin{align} \label{eq:var_pp}
\VV\left[\widehat{\mu}(f)\right] & = \frac{1}{\rho_\YY} \, \int_B f(u)^2\de u
	+ \int_{B^2} (g_\YY(u,v)-1)f(u)f(v)\de u\de v.
\end{align}
If $f$ is non-negative (or non-positive), Equation~\eqref{eq:var_pp} suggests that using a point processes satisfying $g_\YY<1$ makes the variance smaller than the first term which turns out to be the variance under the Poisson case. It is worth noting that the use of a DPP for this task does not require any sign assumption for $f$. Indeed, given the fact that $1-g_\YY(u,v)= |K(u,v)|^2/\rho_\YY^2$ and from Mercer's decomposition~\eqref{eq:Mercer}, we obtain
\begin{equation}\label{eq:Iest_DPP}
\VV\left[\widehat{\mu}(f)\right] = \frac{1}{\rho_\YY} \int_B f(u)^2\de u
	-\frac{1}{\rho_\YY^2} \sum_{j,k\in \NN} \lambda_j \lambda_k   \abs{\int_B f(u) \phi_j(u) \overline{\phi_k}(u) \de u}^2
\end{equation}
and so the second term is always negative.

The use of a general DPP, {{\it i.e. } with random number of points, does not seem to be of great interest: thinking $\rho_\YY=N$ as a number of points, we claim that the rate of convergence remains the same as in the independent case. This claim is based on the previous empirical study done by~\cite{Mazoyeretal20} which tends to show that the empirical variances of Monte-Carlo integral estimates based on quadrature points from a DPP with, for instance, a Gaussian kernel  decrease with rate $N^{-1}$}.

Therefore it seems natural to focus on the subclass of projections DPPs, {\it i.e. } a class for which the number of points is almost surely constant. \cite{BardenetHardy20} have proposed such an approach using an {\it ad-hoc} Orthogonal Polynomial Ensemble with $N$ points. This class is a particular inhomogeneous projection DPP on $B$. As already outlined in the introduction, under the assumption that $f\in C^1(B)$  and that $f$ is compactly supported on some bounded $B^\prime \subset B$, 
it is proved that  a central limit theorem holds for the integral estimator with variance decreasing as $N^{-1-1/d}$. We propose a similar approach, based on a realization of an $(N,d)$-Dirichlet DPP $\XX$.  The $(N,d)$-Dirichlet DPP, detailed in the next section, is a projection DPP based on the Fourier basis. Unlike, the {point process} proposed by~\cite{BardenetHardy20}, this DPP has the advantage to be homogeneous and its projections $\XX_I$ for $I \subseteq \{1,\dots,d\}$ are fully characterized (see Section~\ref{ssec:ProjDirichlet}). Finally, an advantage of our approach is that we do not require that $f$ is continuously differentiable. We only assume that $f$ belongs to some Sobolev space with low regularity parameter, see Section~\ref{sec:MC} and in particular Theorem~\ref{thm:asymp_Iest} for more details.


\section{The $(N,d)$-Dirichlet DPP and its projections} \label{sec:dirichletDPP}
 
\subsection{The $(N,d)$-Dirichlet DPP} 

Let us consider the Fourier basis in $B=\I{0}{1}^d$ defined for any $j\in\ZZ^d$, $x\in B$ by
$\phi_j^{(d)}(x) = \e{2\ii\pi j^\top x}$.
Given a vector of $d$ positive integers $n=(n_i)_{i=1\dots d}$, we construct the following kernel
\begin{equation} \label{eq:dDirichlet}
	K(x,y) = \sum_{j\in E_N} \phi_j^{(d)}(x)\overline{\phi_j^{(d)}(y)} = \sum_{j\in E_N} \e{2\ii \pi j^\top(x-y)}
\end{equation}
where 
\begin{equation} \label{eq:EN}
	E_N = E_1\times \dots \times E_d \quad \text{and for }i=1\dots d, \quad 
	E_i = \{0,1,\dots,n_i-1  \}.
\end{equation}
Thus, $E_N$ is the rectangular subset of $\ZZ^d$ with cardinality $N$ which identifies eigenvalues that are all equal to 1. Due to the invariance by translation of the Fourier basis, the kernel~\eqref{eq:dDirichlet} is  a homogeneous kernel. This construction implies that for any $x,y\in \I{0}{1}^d$, $K(x-y) = \prod_{i=1}^dK_i(x_i-y_i)$ where the $K_i$'s are one-dimensional stationary kernels defined for any  $x_i,y_i\in \I{0}{1}$ by
$K_i(x_i-y_i) = \sum_{j=0}^{n_i-1} \e{2\ii \pi j(x_i-y_i)}  $.
We point out that defining $E_i$ as a block of successive $n_i$ frequencies ({\it e.g. }frequencies centered around 0), would lead to the same DPP \cite[Remark 4, p.48]{Houghetal09}. In particular, if $n_i$ is odd, we could consider $E_i=\{-\floor{n_i/2},\dots,\floor{n_i/2}\}$, which leads to the standard Dirichlet kernel, see {\it e.g. }\cite{Zygmund02}. This justifies the name Dirichlet DPP for this {stochastic process}. 

Such a DPP, which produces almost surely $N=\prod_{i=1}^d n_i$ points in $B$, will be referred to as an $(N,d)$-Dirichlet DPP. We could wonder why we impose $E_N$ to be a rectangular subset of $\ZZ^d$ instead of, for instance, the graded lexicographic order used by~\cite{BardenetHardy20}. As seen in the next section, the rectangular nature of $E_N$ allows us to characterize the distribution of $\XX_I$ for any $I\subseteq \overline d$.

Let us add that, when $d=1$, the kernel $K$  corresponds to the Fourier approximation \cite{Lavancieretal15_1} of the one-dimensional sine-kernel {$\sin(\pi n_i t)/\pi t$ ($t\in(0,1)$)}, which takes its origins in the joint distribution of the eigenvalues (called the Weyl measure) of a unitary matrix. Asymptotic results involving one-dimensional linear functionals from the sine-kernel appear in several papers, see e.g.~\cite{soshnikov:00}. The present paper provides therefore an extension to the $d$-dimensional case and a more thorough treatment of the statistical application to Monte Carlo integration.

{The $(N,d)$-Dirichlet DPP is a homogeneous point process producing exactly $N$ points. Since we do not want to take any advantage from the function to be integrated (and we  could potentially  be interested in estimating several integrals), it is a very natural to use a homogeneous model. We come back to this question in Section~\ref{sec:conclusion}.}

\subsection{Projections of an $(N,d)$-Dirichlet kernel} \label{ssec:ProjDirichlet}

An $(N,d)$-Dirichlet kernel \eqref{eq:dDirichlet} can be written as the product of $d$ one-dimensional kernels. More precisely, for any $I\subset\overline{d}$, by denoting $N_I=\prod_{i\in I} n_i$ and $N_{I^c}=N/N_I$, the $(N,d)$-Dirichlet kernel can always be written as
	\begin{equation}
		\label{eq:KOmegaOmegac}
	K(x-y) = K_I(x_I-y_I) K_{I^c}(x_{I^c}-y_{I^c})
	\end{equation}
where $K_I$ (resp. $K_{I^c}$) is the $(N_I,\iota)$-Dirichlet kernel (resp. $(N_{I^c},d-\iota)$-Dirichlet kernel).
	Projected point processes $\XX_I$ from models with kernels satisfying~\eqref{eq:KOmegaOmegac}
	have been studied and characterized in \cite{Mazoyeretal20}. In particular, for an $(N,d)$-Dirichlet DPP we have the following result. 

\begin{proposition}\label{prop:projDirichlet}
Let $\XX$ be an $(N,d)$-Dirichlet DPP on $B$, let $I \subseteq \{1,\dots,d\}$, then $\XX_I$ is an $(-1/N_{I^c})$-$\DPP$ on $B_I$ with kernel $N_{I^c} K_I$, {\it i.e. } $\XX_I \sim (-1/N_{I^c})$-$\DPP_{B_I}(N_{I^c}K_I)$. In particular, $\XX_I$ has (obviously) $N$ points and $k$-th order intensity
\[
	\rho_{\XX_I}\left(x^{(1)},\dots,x^{(k)}\right)  = \mathrm{det}_{-1/N_{I^c}}\left[N_{I^c} K_I\left(x^{(i)},x^{(j)}\right)\right]_{i,j=1}^k 
\]
for any pairwise distinct $x^{(1)},\dots,x^{(k)} \in B_I$. Its pcf is therefore given, for any pairwise distinct $x,y\in B_I$, by
\begin{equation}
\label{eq:pcf_XOmega}
g_{\XX_I}(x,y) = 1-\frac{\abs{K_I(x,y)}^2}{N N_{I}}.
\end{equation}
\end{proposition}
In the above result, the notation $\det_\alpha$ stands for an $\alpha$ determinant, see e.g.~\cite{ShiraiTakahashi03} for details on such quantities and for general properties of $(\alpha)$-DPPs. 
Proposition~\ref{prop:projDirichlet} therefore proposes a full characterization of the distribution of $\XX_I$. {This result is not directly used in our paper (except in Appendix~\ref{sec:appendix} where we propose an alternative proof to our main result). However,} its main consequence {(for the paper)} is that the pcf of $\XX_I$ is  bounded by 1 (see~\eqref{eq:pcf_XOmega}). Therefore, for any $I$, $\XX_I$ remains in the class of repulsive point patterns.

{We point out that the DPP proposed by~\cite{BardenetHardy20} does not satisfy the general  assumptions of~\cite{Mazoyeretal20}. Therefore for this DPP, the distribution of $\XX_I$ is not explicit and it is unclear whether $\XX_I$ remains repulsive or not.}


\section{Numerical integration with Dirichlet kernel} \label{sec:MC}

\subsection{Objective}

In this section, we study the use of specific DPPs for Monte Carlo integration. To this end, we use notation introduced in Section~\ref{ssec:projectionSPP}. Our objective is to estimate any $\iota$-dimensional integral, for $1\le \iota\le d$, using a Monte Carlo approach and using the same quadrature points. More precisely, let $d\ge 1$, $I \subset \overline d = \{1,\dots,d\}$ with cardinality $\iota=|I|$ and let $f_I: B_I \to \RR$ be a measurable function on $B_I = \I{0}{1}^\iota$, such that $f_I\in L^2(B_I)$. More assumptions on $f_I$ will be given later.
We intend to estimate
\[
	\mu(f_I) = \int_{B_I} f_I (u) \de u
\]
using the projection onto $B_I$ of $\XX$ an $(N,d)$-Dirichlet {DPP} on $B$. In particular, we estimate $\mu(f_I)$ by 
\begin{equation} \label{eq:def_Iest}
	\widehat{\mu}_N( f_I) = \frac1N \sum_{u \in \XX_I} f_I(u) = \frac1N \sum_{j=1}^N f_I ((u_j)_I)
\end{equation}
where $\XX= \{u_1,\dots,u_N\}$ is an $(N,d)$-Dirichlet DPP on $B$ and where we remind the notation $(u_j)_I=P_I u_j$ for any $u_j\in B$.
In the following, we study asymptotic properties for $\widehat{\mu}_N(f_I)$. 

In this paper, we have chosen to focus on integrals on $\I{0}{1}^\iota$ for simplicity. It can straightforwardly be extended to rectangles. Indeed, let $a=(a_i)_{i=1,\dots,d}, b=(b_i)_{i=1,\dots,d} \in \RR^d$ such that $a_i<b_i$, $i=1,\dots,d$, let $R = \I{a_1}{b_1}\times \dots \times \I{a_d}{b_d}$. {Denoting $R_I=\prod_{i\in I} \I{a_i}{b_i} $,} an estimate of $\int_{R_I} f_I(u) \de u$ where now $f_I \in L^1(R_I)$ is simply given by
\begin{equation}
	\label{eq:Iest_rectangle}
	\widehat \mu_N(f_I) = 
\prod_{i\in I} (b_i-a_i) \; \left\{ \frac{1}N \sum_{i=1}^N  f_I \left( 
a_I + (b_I-a_I)(u_j)_I 
\right) \right\}.
	\end{equation}

We introduce two additional fundamental pieces of notation induced by the choice of the Fourier basis and used in our results. Let $f_I \in L^2(B_I)$. The notation $\widehat f_I(j)$ for $j\in \ZZ^\iota$ stands for the $j$th Fourier coefficient, {\it i.e. }
\begin{equation}\label{eq:fhatj}
	\widehat f_I(j) = \int_{[0,1]^\iota} f_I(u) \e{-2\ii\pi j^\top u} \mathrm d u.
\end{equation}
 Finally, we define the space $\mathcal H^s(B_I)$ as the isotropic (with respect to the sup norm) Sobolev space with index $s\ge 0$ of square integrable periodic functions by
 \begin{equation}
 \label{eq:defHs}
 \mathcal H^s(B_I) = \left\{
f_I \in L^2_{\mathrm{per}}(B_I), \, : \sum_{j \in \ZZ^\iota} \left( 1+ \|j\|_\infty\right)^{2s} \, |\widehat f_{{I}}(j)|^2<\infty
 \right\}
 \end{equation}
 where $L^2_{\mathrm{per}}(B_I)$ is the set of square integrable periodic functions. By $f_I$ periodic, we mean that {for any $i\in I$
 \[
f_I(x_1,...,x_{i-1}, 0, x_{i+1}, ..., x_\iota) = f_I(x_1,...,x_{i-1}, 1, x_{i+1}, ..., x_\iota).
\]
Note that, if $f_I$ does not satisfy this condition, we can for example consider the function 
$g_I(x_1,\dots,x_\iota) = f_I(\abs{x_1},\dots,\abs{x_\iota})$ which will satisfy this periodic condition on $[-1,1]^\iota$, and $\widehat{\mu}_N(f_I) = 2^{-\iota}\widehat{\mu}_N(g_I)$.
}

{In the introduction, we mention that several Monte-Carlo integration methods (stratified-based methods, scrambled nets, {\it etc}) require some (strong) regularity assumptions of the integrands. In connection with the Sobolev space $\mathcal H^s(B_I)$, we remind that (see {\it e.g. }\cite{robinson}) for any $\varepsilon>0$
\begin{equation}\label{eq:sobol}
\widetilde{\mathcal H}^{3/2+\varepsilon}(B_I) \; \subset \; C^1_{\mathrm{per}}(B_I) \; \subset \; \mathcal H^1(B_I).
\end{equation}
where 
\begin{align*}
 C^1_{\mathrm{per}}(B_I) &= \left\{ f_I \; \text{periodic}, f_I \in C^1(B_I)\right\}	\\
 \widetilde{\mathcal H}^{3/2+\varepsilon}(B_I) &= \left\{ f_I \in L^2_{\mathrm{per}}(B_I):  \widehat f_I(j) = \mathcal O(\|j\|_\infty^{-2-\varepsilon}) 
 \right\}, \quad \varepsilon>0. 
 \end{align*}
 (Of course, $\mathcal H^{3/2+\varepsilon}(B_I)$ and $\widetilde{\mathcal H}^{3/2+\varepsilon}(B_I)$ are very similar). In the same way, for any function $f_I$ periodic and $d$ times continuously differentiable, belongs to $\mathcal H^d(B_I)$.
 }

Our main results expressed by Theorem~\ref{thm:asymp_Iest} and Corollary~\ref{thm:asymp_IestProj}, have the interest to focus on functions $f_I \in \mathcal H^s(B_I)$ with small regularity parameter $s>0$, thus, according to~\eqref{eq:sobol}, potentially to non-differentiable functions. In particular, we show that a CLT {holds} as soon as $s>1/2$.

Finally, let us justify why we focus on periodic functions on $B_I$. For a non periodic function which is at least twice continuously differentiable it is a known fact that $|\widehat f_I(j)| = \mathcal O(\|j\|_\infty^{-1})$. So, the summability condition in~\eqref{eq:defHs} for such a smooth function would be fulfilled only for $s<1/2$, {a situation where no CLT is available (even if $f_I$ is infinitely differentiable).}

\subsection{Case $\iota= d$} \label{ssec:Iest}

We first consider the $d$-dimensional case. Let $f\in L^2_{\mathrm{per}}(B)$ with $B=\I{0}{1}^d$. The first result shows how the variance given in \eqref{eq:Iest_DPP} relates to the Fourier coefficients of $f$ in the case of an $(N,d)$-Dirichlet DPP. 

\begin{proposition} \label{prop:reductionvariance}
 Let $\XX$ be an $(N,d)$-Dirichlet DPP and $f\in L_{\mathrm{per}}^2(B)$. For any $j\in \ZZ^d$, we let $c_j(f)$ denote the $j$th Fourier coefficient of $f$. Then, $\widehat{\mu}_N(f)$ given by~\eqref{eq:def_Iest} is an unbiased estimator of $\mu(f)=\int_B f(u) \de u$ with variance given by
\begin{align} 
\VV\left[\widehat{\mu}_N(f)\right] &= \frac1N \sum_{j \in \ZZ^d} \abs{\widehat f(j)}^2 
	- \frac1{N^2} \sum_{j,k \in E_N}  \abs{\widehat f(j-k)}^2  \label{eq:var1}\\
& = \frac{1}{N}\sum_{j\in \ZZ^d} \abs{\widehat f(j)}^2 - \frac{1}{N^2}\sum_{j\in F_N} \left[\prod_{i=1}^d\left(n_i-\abs{j_i}\right)\right]\abs{	\widehat f(j)}^2 \label{eq:var2}
\end{align}
where $F_N = \{j \in \ZZ^d: |j_i| \le n_i-1, \; i=1,\dots,d\}$.
\end{proposition}

This simple form of the variance of $\widehat{\mu}_N(f)$ invites us to study its asymptotic behavior as $N \to \infty$. Given the fact that $N= \prod_i n_i$, we require a specific asymptotic. For $i=1,\dots ,d$, we assume that $(N_\nu)_{\nu \ge 1}$ and $(n_{i,\nu})_{\nu \ge 1}$ are integer sequences  indexed by some $\nu\ge 1$. We assume that each sequence $n_{i,\nu}$ tends to $\infty$ as $\nu \to \infty$ and that there exist $\kappa_i>0$ for $i=1,\dots,d$ such that
\begin{equation} \label{eq:cont_asymp}
\lim_{\nu\to\infty} n_{i,\nu} N_\nu^{-1/d} = \kappa_i.
\end{equation}
For the sake of conciseness, we skip the dependence in $\nu$. Similarly, when we write $N\to \infty$, we implicitly assume~\eqref{eq:cont_asymp}.

We can now obtain the following asymptotic behavior of the variance of $\widehat{\mu}_N(f)$ and for some values of $s$ a central limit theorem. Regarding this asymptotic normality, when $d=1$, as mentioned in Section~\ref{sec:dirichletDPP}, $\XX$ corresponds, up to a normalization, to the joint distribution of the eigenvalues of a unitary matrix. Linear statistics for such a DPP have been deeply studied in the literature, see e.g.~\cite{soshnikov:00} and the references therein. When $d>1$, Theorem~\ref{thm:asymp_Iest} (iii) is therefore original, and relies upon~\cite[Theorem~1]{soshnikov:02}.

\begin{thm} \label{thm:asymp_Iest}
Consider the asymptotic framework~\eqref{eq:cont_asymp} and assume that $f \in \mathcal H^s(B)$. Then, we have the following statements.

(i) If $s\in(0{,}1/2)$, then as $N\to \infty$
\[
	\VV \left[\widehat{\mu}_N(f)\right] = \mathcal O \left( N^{-1-\frac{2s}d}\right).
\]

(ii) If $s\geq 1/2$ for $d=1$ or $s>1/2$ for $d>1$, then
\begin{equation} \label{eq:var_asymp}
\lim_{N\to\infty} N^{1+1/d} \VV\left[\widehat{\mu}_N(f)\right] = \sigma^2(f) = \sum_{j\in\ZZ^d}\left(\sum_{i=1}^d \frac{\abs{j_i}}{\kappa_i}\right) \abs{\widehat f(j)}^2
\end{equation}
where $\widehat f(j)$ is given by~\eqref{eq:fhatj}.

(iii) {If $s\geq 1/2$ for $d=1$ or $s>1/2$ for $d>1$ and, if in addition,} $\|f\|_\infty<\infty$, then as $N\to \infty$,
\begin{equation}
	\label{eq:clt}
	\sqrt{N^{1+1/d} } \left( \; \widehat{\mu}_N (f) - \mu(f) \;  \right) \to {N}(0,\sigma^2(f))
\end{equation}
in distribution.

\end{thm}

Let us rephrase Theorem~\ref{thm:asymp_Iest} (i): if $f \in \mathcal H^s(B)$ for some $s>0$, then necessarily $\VV[\widehat{\mu}_N(f)]=o(N^{-1})$. That is, the variance of the estimator  proposed decreases to 0 faster than the standard Monte Carlo estimator,  as soon as $\abs{\widehat f(j)} = \mathcal O (\|j\|_\infty^{-\varepsilon})$ for some $\varepsilon>0$, which is a very weak assumption.

Theorem~\ref{thm:asymp_Iest} (ii)  shows that the variance $\sigma^2(f)$ can be {approximated} by $\widehat \sigma_N^2(f)$  given by
\begin{equation}  \label{eq:est_asymp_const}
	\widehat \sigma_N^2(f)	 = N^{1/d} \sum_{j\in F_N} \left(\sum_{i=1}^d\frac{\abs{j_i}}{n_i}\right) \abs{\widehat f(j)}^2
\end{equation}
which has the interest to avoid the constants $\kappa_i$ {defined in \eqref{eq:cont_asymp}}. {We point out that \eqref{eq:est_asymp_const} does not provide an estimate of $\sigma^2(f)$. We come back to this question in Section~\ref{sec:conclusion}.}

\subsection{Discussion on $\sigma^2(f)$}

{In this section, we would like to discuss how the value of $\sigma^2(f)$ varies with  $d$ and how this variance compares to standard Monte-Carlo method and the stratified-based Monte-Carlo method (which are important alternatives for which a central limit theorem is available). For the sake of simplicity, we set in this section $\kappa_i=1$. For $f\in \mathcal{H}^s([0,1]^d)$ with $s\ge 0$, we  denote by 
\[
	\| f\|_{\mathcal{H}^s([0,1]^d)}^2 = \sum_{j\in \ZZ^d} \|j\|^{2s}_{2s} |\hat f(j)|^2
\]
where $\|j\|^{p}_p=\sum_{l=1}^d |j_l|^p$. With such notation and recalling that $\kappa_i=1$ here we get $\sigma^2(f)=\|f\|_{\mathcal H^{1/2}([0,1]^d)}^2$. The standard Monte-Carlo approach (resp. the stratified-based Monte-Carlo method, see \cite{owen:book:13}) assumes that $f\in L^2([0,1]^d)$ (resp. $f$ is continuously differentiable on $[0,1]^d$). The corresponding estimator has variance descreasing as $N^{-1}$ (resp. $N^{-1-2/d}$) with asymptotic constants given by
\begin{align*}
\sigma^2_{\mathrm{std}}(f)	&= \|f\|^2_{L^2([0,1]^d)}- \mu(f)^2 = \|f-\mu(f)\|^2_{\mathcal H^0([0,1]^d)}\\ 
\sigma^2_{\mathrm{strat}}(f)	&= \| \;\|\nabla f\|_2 \; \|^2_{L^2([0,1]^d)} = 4\pi^2 \|f\|_{\mathcal H^1([0,1]^d)}^2.
\end{align*}
The latter expression ensues from Parseval's identity and standard expression of Fourier coefficients for continuously differentiable functions.}

{To compare these three constants in terms of $d$, we consider two particular cases for $f$:
\begin{itemize}
\item[(i)] $f$ as a product of one-dimensional functions: $\exists f_0\in L^2([0,1])$ such that
\begin{equation}\label{eq:f0prod}
f(x) = \prod_{i=1}^d f_0(x_i), \qquad x \in [0,1]^d.	
\end{equation}
\item[(ii)] $f$ as a sum of one-dimensional functions: $\exists f_0\in L^2([0,1])$ such that
\begin{equation}\label{eq:f0sum}
f(x) = \sum_{i=1}^d f_0(x_i), \qquad x \in [0,1]^d.	
\end{equation}
\end{itemize}
We are now able to present the following result.}

{\begin{proposition}\label{prop:sigma2} ${ }$\\
(i) Let $f$ be defined as~\eqref{eq:f0prod} and satisfying the appropriate assumptions of the considered method, then
\begin{align*}
\sigma^2(f) &= d \,\|f_0\|^2_{\mathcal H^{1/2}([0,1])} \, \|f_0\|^{2d-2}_{\mathcal H^0([0,1])}\\
\sigma^2_{\mathrm{std}}(f) &= \|f_0\|^{2d}_{\mathcal H^0([0,1])} - \mu(f_0)^{2d} \\
\sigma^2_{\mathrm{strat}}(f) &= 4\pi^2 d\, \|f_0\|^2_{\mathcal H^{1}([0,1])} \, \|f_0\|^{2d-2}_{\mathcal H^0([0,1])}.\\	
\end{align*}
(ii) Let $f$ be defined as~\eqref{eq:f0sum} and satisfying the appropriate assumptions of the considered method, then
\begin{align*}
\sigma^2(f) &= d \,\sigma^2(f_0) = d \,\|f_0\|^2_{\mathcal H^{1/2}([0,1])} \\
\sigma^2_{\mathrm{std}}(f) &= d \sigma^2_{\mathrm{std}}(f_0)=  d \, \|f_0-\mu(f_0)\|^2_{\mathcal H^0([0,1])}\\ 
\sigma^2_{\mathrm{strat}}(f) &= d \, \sigma^2_{\mathrm{strat}}(f_0) = 
4\pi^2 d\, \|f_0\|^2_{\mathcal H^{1}([0,1])}. \\
\end{align*}	
\end{proposition}}

{Proposition~\ref{prop:sigma2} shows that when $f$ is given as the product (resp. sum) of one-dimensional functions, then $\sigma^2(f)$ varies exponentially (resp. linearly) with the dimension $d$. This result points out that the constants for the standard Monte-Carlo and stratified-based Monte-Carlo methods vary along the same lines.}

\subsection{Case $I\subset \overline{d}$} \label{ssec:Iest_proj}

We now consider the situation where we estimate $\mu(f_I)$ (on $B_I$) based on 
$\{u_1,\dots,u_N\}$ which is an $(N,d)$-Dirichlet DPP on $B$. 
In this section, we naturally assume that $d>1$. The interest of the contruction of {the Dirichlet-DPP} is revealed by Corollary~\ref{thm:asymp_IestProj} which, briefly, states that Theorem~\ref{thm:asymp_Iest} can be applied to functions of the form $f_I^\uparrow(x) = f_I(x_I)\ind{x_{I^c}\in B_{I^c}}$, $x\in B$.

\begin{corollary} \label{thm:asymp_IestProj} \label{THM:ASYMP_IESTPROJ}
Let $d>1$ and $I\subset \{1,\dots,d\}$  with cardinality $\iota>0$. Consider the asymptotic framework~\eqref{eq:cont_asymp} \and assume that $f_I\in \mathcal H^s(B_I)$, then  the following statements hold.\\

(i) If $s\in (0{,}1/2)$, then as $N\to \infty$
\[
	\VV \left[\widehat{\mu}_N(f_I)\right] = \mathcal O \left( N^{-1-\frac{2s}d}\right).
\]

(ii) If $s>1/2$, then
\begin{equation} \label{eq:var_asymp_Proj}
\lim_{N\to\infty} N^{1+1/d} \VV\left[\widehat{\mu}_N(f_I)\right] = \sigma^2(f_I) 
\end{equation}
where 
\begin{equation}
	\label{eq:sigmafOmega}
	\sigma^2(f_I) = \sum_{j\in\ZZ^\iota}\left(\sum_{i\in I} \frac{\abs{j_i}}{\kappa_i}\right) \abs{\widehat f_I(j)}^2
\end{equation}
and where $\widehat f_I(j)$ is given by~\eqref{eq:fhatj}.

(iii) {If $s>1/2$ and }  $f_I$ is bounded, then as $N \to \infty$
\begin{equation}
	\label{eq:cltProj}
	\sqrt{N^{1+1/d} } \left( \; \widehat{\mu}_N (f_I) - \mu(f_I) \;  \right) \to N\left(0,\sigma^2(f_I)\right)
\end{equation}
in distribution.
\end{corollary}

The asymptotic constant $\sigma^2(f_I)$ can still be {approximated} by \eqref{eq:est_asymp_const}. The proof of this result is a straightforward  consequence of Theorem~\ref{thm:asymp_Iest}. Another approach using the fact that $\XX_I$ is distributed as an $\alpha$-DPP is proposed in Appendix~\ref{sec:appendix}.

{To rephrase Corollary~\ref{thm:asymp_IestProj}, we can estimate $\mu(f_I)$ for any $\iota$-dimensional function $f_I$ with the rate of convergence  $\sqrt{N^{1+1/d}}$ (if $f_I \in \mathcal H^s(B_I)$ with $s>1/2$), which corresponds to the rate of convergence from the dimension where the points were generated. This is  the price to pay to be able to estimate any function in any dimension. Of course, if one knows beforehand that we would like to estimate only, say one-dimensional integrals, one could use the decomposition $N=N\times 1\times \dots\times 1$ for the set $E_N$ of eigenvalues, see~\eqref{eq:EN}. Using such a DPP (in dimension $d>1$), we claim that any one-dimensional integral could be estimated with a rate of convergence $\sqrt{N^2}$ which would be a gain with respect to $\sqrt{N^{1+1/d}}$. However, the resulting DPP would be catastrophic to estimate $\mu(f_I)$ for any $I$ such that $\iota>1$.}


\subsection{Multivariate central limit theorem}

We can combine Theorem~\ref{thm:asymp_Iest} and Corollary~\ref{thm:asymp_IestProj} to obtain a multivariate version of the central limit theorem. 

\begin{corollary} \label{cor:multivariate}
Let $p\ge 1$. For any $\ell=1,\dots,p$, let $I_\ell\subseteq \{1,\dots,d\}$ with $\iota_\ell=|I_\ell|$ and assume that $f_{I_\ell} \in \mathcal H^s(B_{I_\ell})$ with $s\ge 1/2$ if $d=1$ or $s>1/2$ if $d>1$.
Let $\widehat{\mu}_{N,p} = \left( \widehat{\mu}_N(f_{I_1}), \dots,\widehat{\mu}_N(f_{I_p}) \right)^\top$ and $\mu_{p} = \left(\mu(f_{I_1},\dots,\mu(f_{I_p})) \right)^\top$. Then, under the asymptotic framework~\eqref{eq:cont_asymp}, $\widehat{\mu}_{N,p}$ is an unbiased estimator of $\mu_p$ and as $N\to \infty$
\[
	\sqrt{N^{1+1/d}} \left( \widehat{\mu}_{N,p} - \mu_p\right) \to N(0,\boldsymbol\Sigma_p)
\]
in distribution, where $\boldsymbol\Sigma_p$ is the $(p,p)$ Hermitian matrix with entries
\begin{equation}\label{eq:defSigmap}
\left(\boldsymbol\Sigma_p \right)_{\ell \ell^\prime}	\; = \; \sum_{j \in \mathbb Z^d} \left(\sum_{i=1}^d \frac{|j_i|}{\kappa_i}\right) \widehat f_{I_\ell}^\uparrow(j) \overline{\widehat f_{I_{\ell^\prime}}^\uparrow(j)}
\end{equation}
where $f_{I_\ell}^\uparrow(x) = f_{I_\ell}(x_I) \ind{x_{I_\ell^c}\in B_{I_\ell^c}}$, $x\in B$.
\end{corollary}

\section{Simulation study} \label{sec:exp}

We propose now a simulation study to illustrate the results. We first consider the setting of Theorem~\ref{thm:asymp_Iest}~(ii)-(iii) and Corollary~\ref{thm:asymp_IestProj}~(ii)-(iii), {\it i.e. } for the case $s>1/2$ (when $d>1$) and and for the situation $s<1/2$.

\subsection{Case $s>1/2$, illustration of Theorem~\ref{thm:asymp_Iest}~(ii)-(iii)} \label{ssec:sim1}

We consider three different functions with different regularity properties:
\begin{itemize}
	\item Bump function 
	\begin{equation}\label{eq:fbump}
		f_\text{bump}(x) = \prod_{i=1}^d \frac{\varphi_{\text{bump}}(x_i)}{\int_0^1 \varphi_{\text{bump}}(t)\de t}, \qquad \varphi_{\text{bump}}(t) = 
	\exp \left( - \frac{0.1}{t(1-t)}\right).
	\end{equation}
	\item {Sum of cosines}
	\begin{equation}\label{eq:fmixcos}
		f_\text{mixcos}(x) = \frac{1}{d}\sum_{i=1}^d \frac{\varphi_{\text{mc}}(x_i)}{\int_0^1 \varphi_{\text{mc}}(t)\de t}, \quad
	\varphi_{\text{mc}}(t) = 0.1\abs{\cos(5\pi (t-1/2))}+(t-1/2)^2.
	\end{equation}
	\item {Product of cosines}
	\begin{equation}\label{eq:fmixcosprod}
		f_\text{mixcosprod}(x) = \prod_{i=1}^d \frac{\varphi_{\text{mc}}(x_i)}{\int_0^1 \varphi_{\text{mc}}(t)\de t}
	\end{equation}
	\item Normalized $L^\gamma$-norm: let $\gamma>0$.
	\begin{equation}\label{eq:fgamma}
	f_{\gamma}(x) = \frac{1}{d}\sum_{i=1}^d \frac{\varphi_{\gamma}(x_i)}{\int_0^{{1}} \varphi_\gamma(t)\de t}, \quad
\varphi_\gamma(t) = |t-1/2|^\gamma.	
	\end{equation}
\end{itemize}
It is worth mentioning that $f_\text{bump}$ is infinitely continuously differentiable, so $f_\text{bump} \in \mathcal H^s$, for any $s>0$. The function $f_\text{mixcos}$ is a non-differentiable (with $4^d$ singularity points) which satisfies $\widehat f_\text{mixcos}(j) = \mathcal O(\|j\|_\infty^{-2})$, so $f_{\text{mixcos}} \in \mathcal H^s(B)$ for any $s<3/2$. {$f_\text{mixcosprod}$ is also a non-differentiable function, which satisfies $\widehat f_\text{mixcosprod}(j) = \mathcal O(\prod_{i=1}^d \|j_i\|^{-2})$, whereby it can be deduced that $f_{\text{mixcosprod}} \in \mathcal H^s(B)$ for any $s<3/2$}. Finally, $f_\gamma$ is also a non-differentiable square integrable periodic function, which satisfies $\widehat f_\gamma(j) = \mathcal O(\|j\|_\infty^{-1-\gamma})$. Hence, $f_\gamma \in \mathcal H^s(B)$ for any $s<1/2+\gamma$. In the following, we consider the cases $\gamma=0.25$ and $\gamma=0.75$.
These {five test} functions are depicted for $d=2$ in Figure~\ref{fig:plot_func}.

\begin{figure} 
\centering
\renewcommand{\arraystretch}{0}
\renewcommand{\tabcolsep}{0pt}
\vspace*{-3.5cm}
\begin{tabular}{M{7cm}M{7cm}}
	\multicolumn{2}{c}{\includegraphics[width=0.5\textwidth]{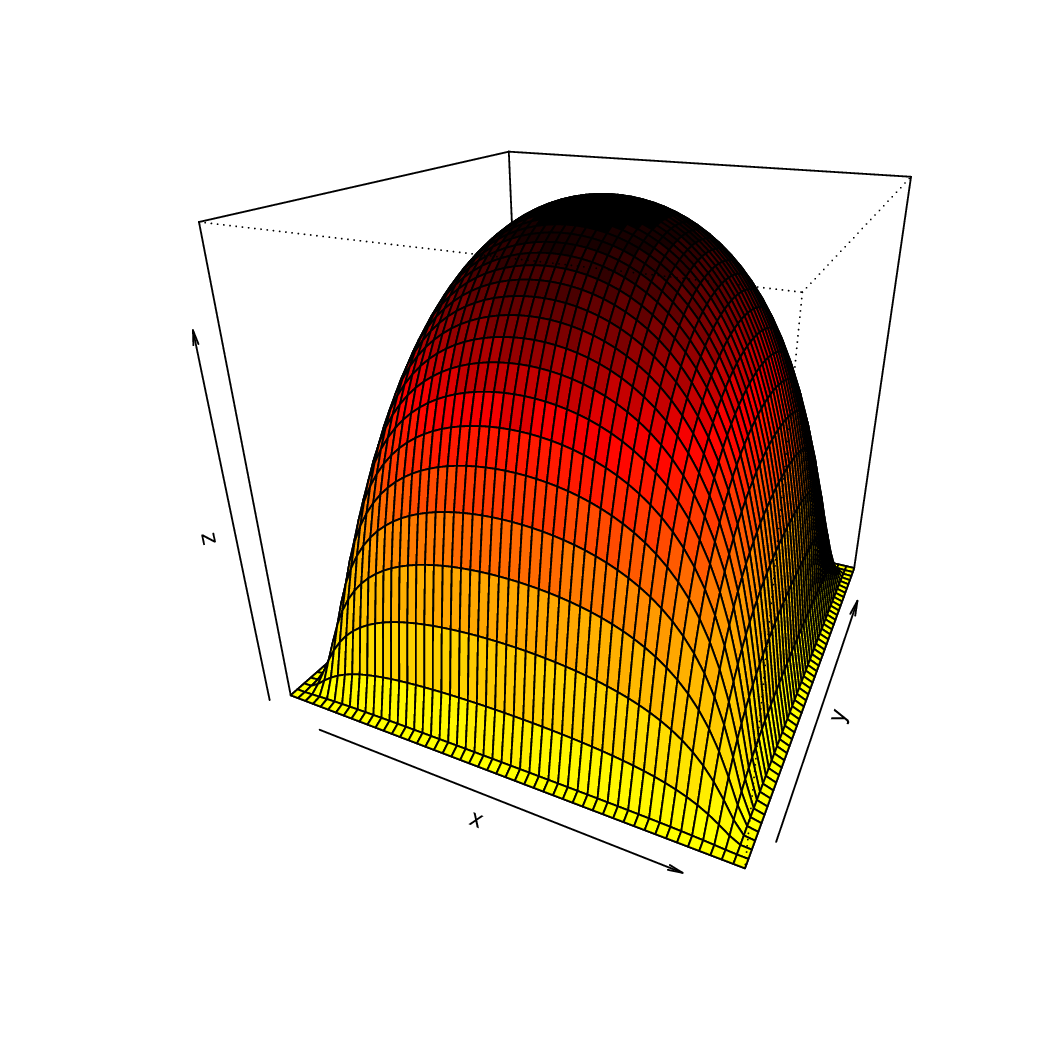}} \tabularnewline
	\multicolumn{2}{c}{$f_\text{bump}(x)$} \tabularnewline
	 \includegraphics[width=0.5\textwidth]{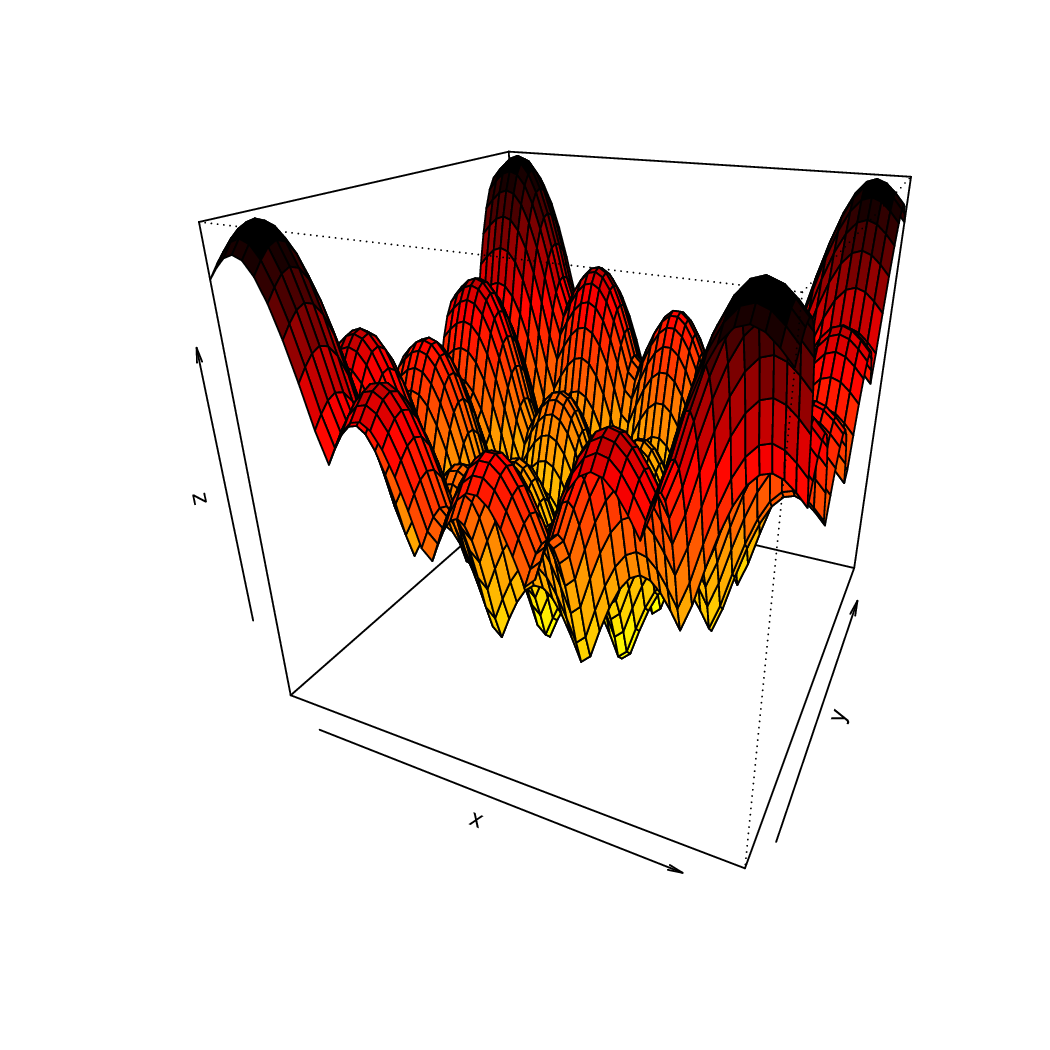} & \includegraphics[width=0.5\textwidth]{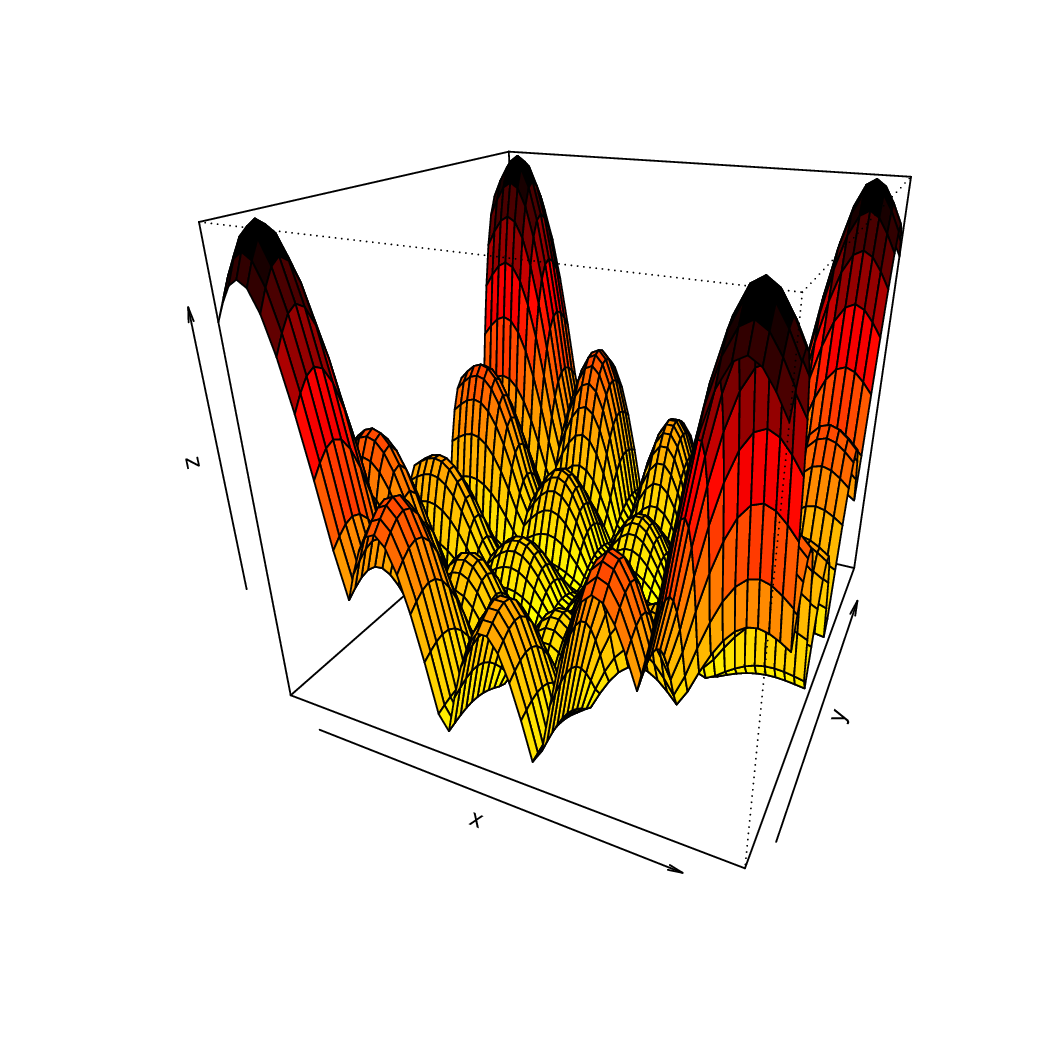} \tabularnewline
	 $f_\text{mixcos}(x)$ & $f_\text{mixcosprod}(x)$ \tabularnewline
	\includegraphics[width=0.5\textwidth]{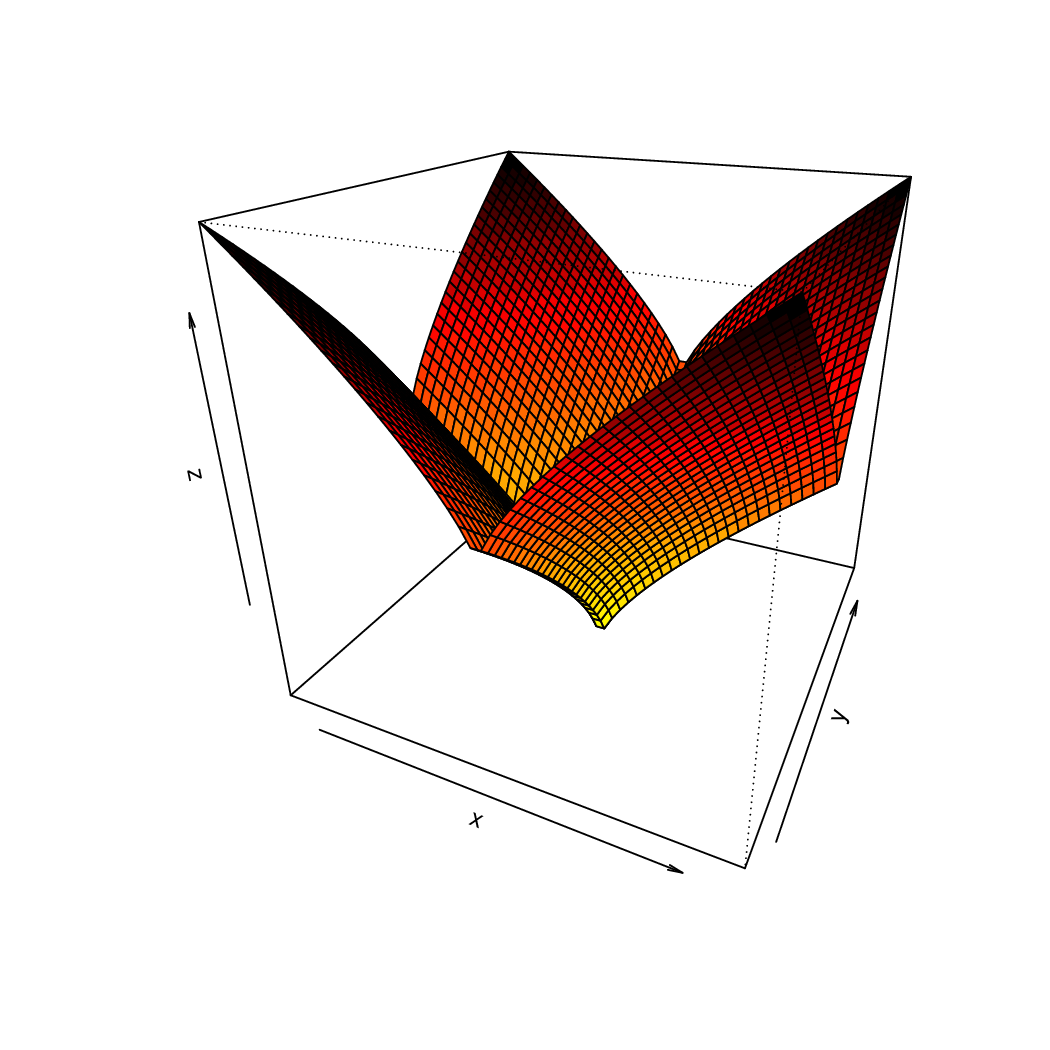} & \includegraphics[width=0.5\textwidth]{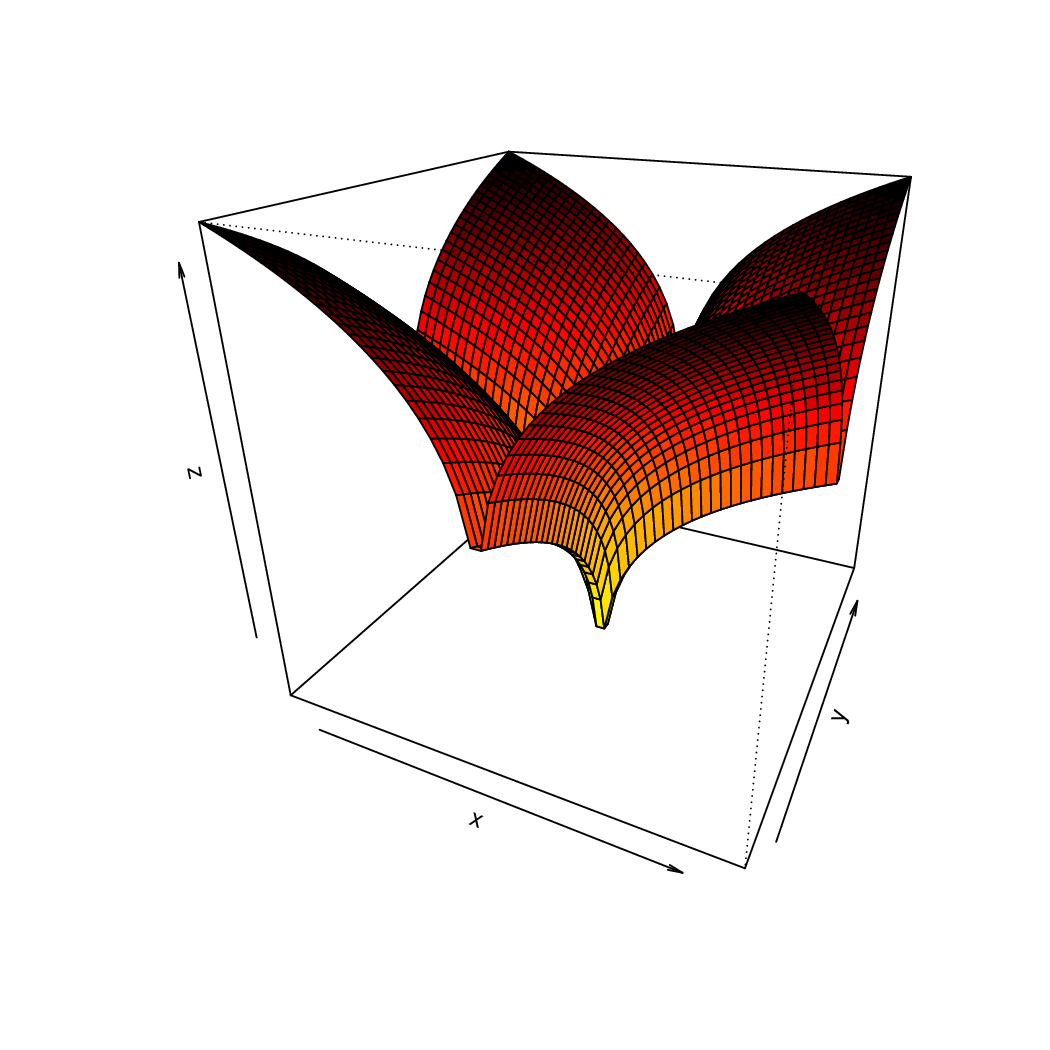} \tabularnewline
$f_\gamma(x)$ for $\gamma=0.75$ &  $f_\gamma(x)$  for $\gamma=0.25$
\end{tabular}
\caption{Test functions considered in the simulation study, given by~\eqref{eq:fbump}-\eqref{eq:fgamma} depicted in dimension $d=2$.}
\label{fig:plot_func}
\end{figure}

We perform the following experiment. For $N=50,100,150,\dots,\linebreak500,600,\dots,1000$ ({\it i.e. } 15 values for $N$) and for $d=1,\dots,6{,10}$, we generate 2500 realizations of the $(N,d)$-Dirichlet DPP. The factorization $N=\prod_{i=1}^d n_i$ is set such that the fluctuation of the $n_i$'s is minimized. For example, for $N=100$ and $d=2$, we set $n=(10,10)$ while when $d=6$ we choose $n=(5,5,2,2,1,1)$.

We use the simulation algorithm provided by \cite{Houghetal06}. Basically, it relies upon a Gram-Schmidt orthogonalization of $N$ vectors with dimension $N$, with a cost of order $N^3$, and a {costly} rejection sampling step {for the simulation of each point location.}
Sampling DPPs with large $N$ or when $d>4$ is very time consuming using the standard \texttt R package \texttt{spatstat} \cite{baddeley:rubak:turner:15}. Therefore, we have reimplemented an \texttt R package based on C++ functions  (available at \url{https://github.com/AdriMaz/rcdpp/}). We perform our experiments in very reasonable computing time, with a {basic laptop} (2,3 GHz Intel Core i5 processor, 8 Go (2133 MHz DDR4) of RAM). For example, sampling a DPP with $N=1000$ in dimension 6 can be performed {within a reasonable time (approximately one minute)}.

For each $d$, each function and each replication of the point pattern, we evaluate the estimator $\widehat \mu_N(f_I)$ given by~\eqref{eq:def_Iest} with $I=\overline{d}$. To visualize the rate of convergence of the variance, we perform a linear regression of the logarithm of empirical variances in terms of $\log(N)$. According to Theorem~\ref{thm:asymp_Iest} (ii), the expected slope is $-1-1/d$. For each test function, $d$ and $N$, we also test the normality of estimates using the Shapiro-Wilk test {\cite{shapiro:wilk:1965}}, after adjusting the $p$-values using Holm procedure {\cite{holm:1979}} for each function and each $d$. {Results are reported in Figure~\ref{fig:ResMCComp}. The fourth columns of the summary tables expose the usual Student-t confidence interval for the slopes. The values of $N$ for which the normality assumptions has been rejected, {\it i.e. } for which adjusted $p$-value is smaller than 0.05, are represented by crosses instead of regular dots. We still made the arbitrary choice to keep them when performing the regression lines. We have also arbitrarily translated the curves to focus the interpretation on the slopes. Thus in Figures \ref{fig:ResMCComp}-\ref{fig:ResMCCompWithProj6DBH} (as well as figures presented in Appendix~\ref{app:simulations}) the $y$-axis has no meaning.}

{Most of results} are in a clear agreement with our theoretical result. {For small values of $d$,} the asymptotic normality is not rejected even for small sample size $N$, and the slope of the logarithm of empirical variances is very close to $-1-1/d$.} 

{For $d=10$, the theoretical results are hardly recovered. This is related to the values of $N$ considered: the values of the $n_i$'s are very small (several of them are actually equal to 1).}

{The structure of the integrands affects the quality of empirical results: for functions $f_\text{bump}$ and $f_\text{mixcosprod}$, normality requires higher values of $N$. The expected slope is also not always included in the confidence intervals. However, this might be related to usual observations when estimating integral of functions defined as \eqref{eq:f0prod}, as illustrated in Section~\ref{sec:comparisons}
.}

\begin{figure} 
	\vspace*{-2.5cm}
	\includegraphics[scale=.15]{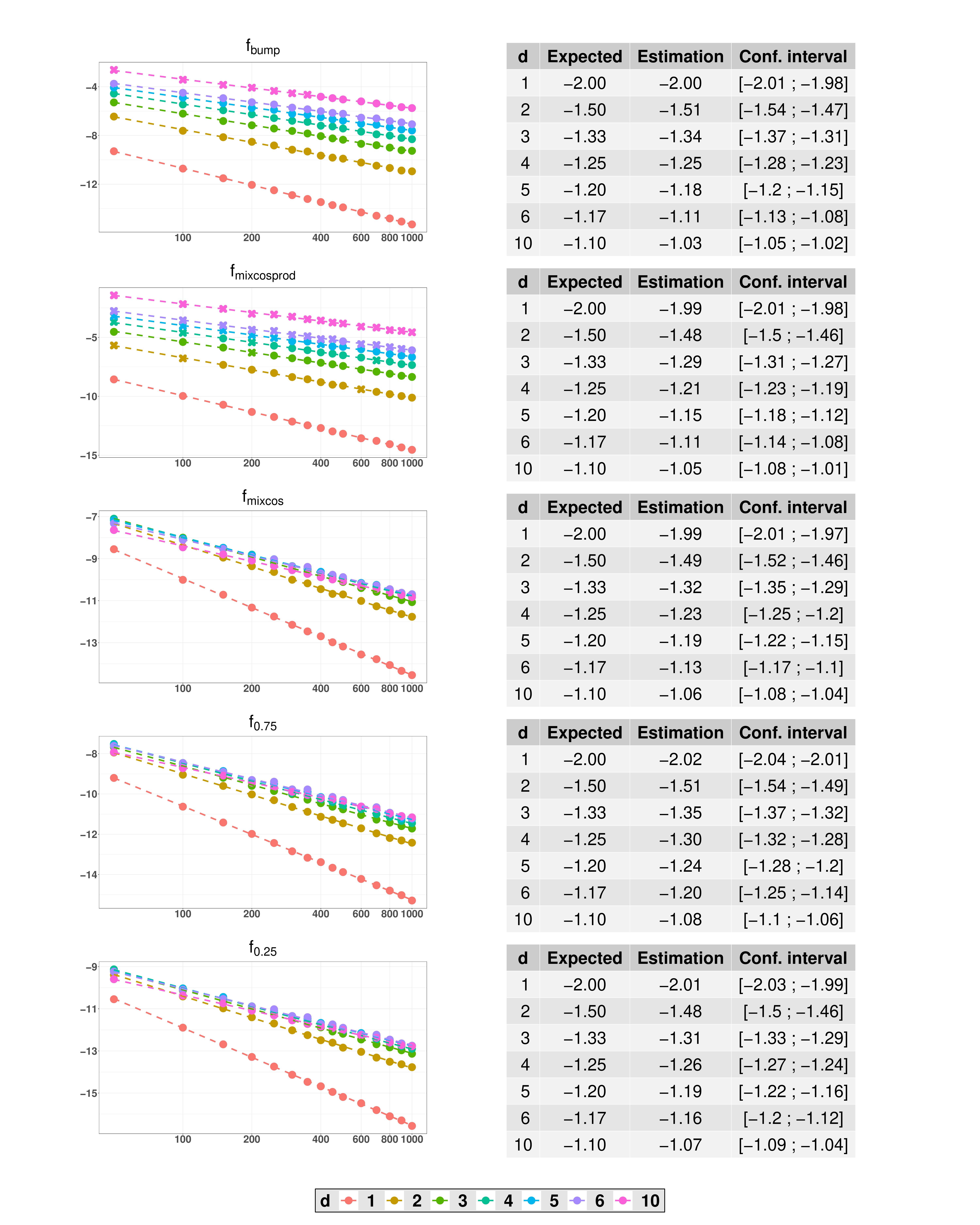}
	\caption{Summary of experiments in which integrals of $d$-dimensional functions are estimated using an $(N,d)$-Dirichlet DPP.  A $\bullet$ (resp. $\times$) indicates that the adjusted p-value of the Shapiro-Wilk test is not smaller (resp. smaller) than $5\%$}
	\label{fig:ResMCComp}
\end{figure}

\subsection{Case $s>1/2$, illustration of Corollary~\ref{thm:asymp_IestProj}~(ii)-(iii)}

\begin{figure} 
	\vspace*{-2.5cm}
	\includegraphics[scale=.16]{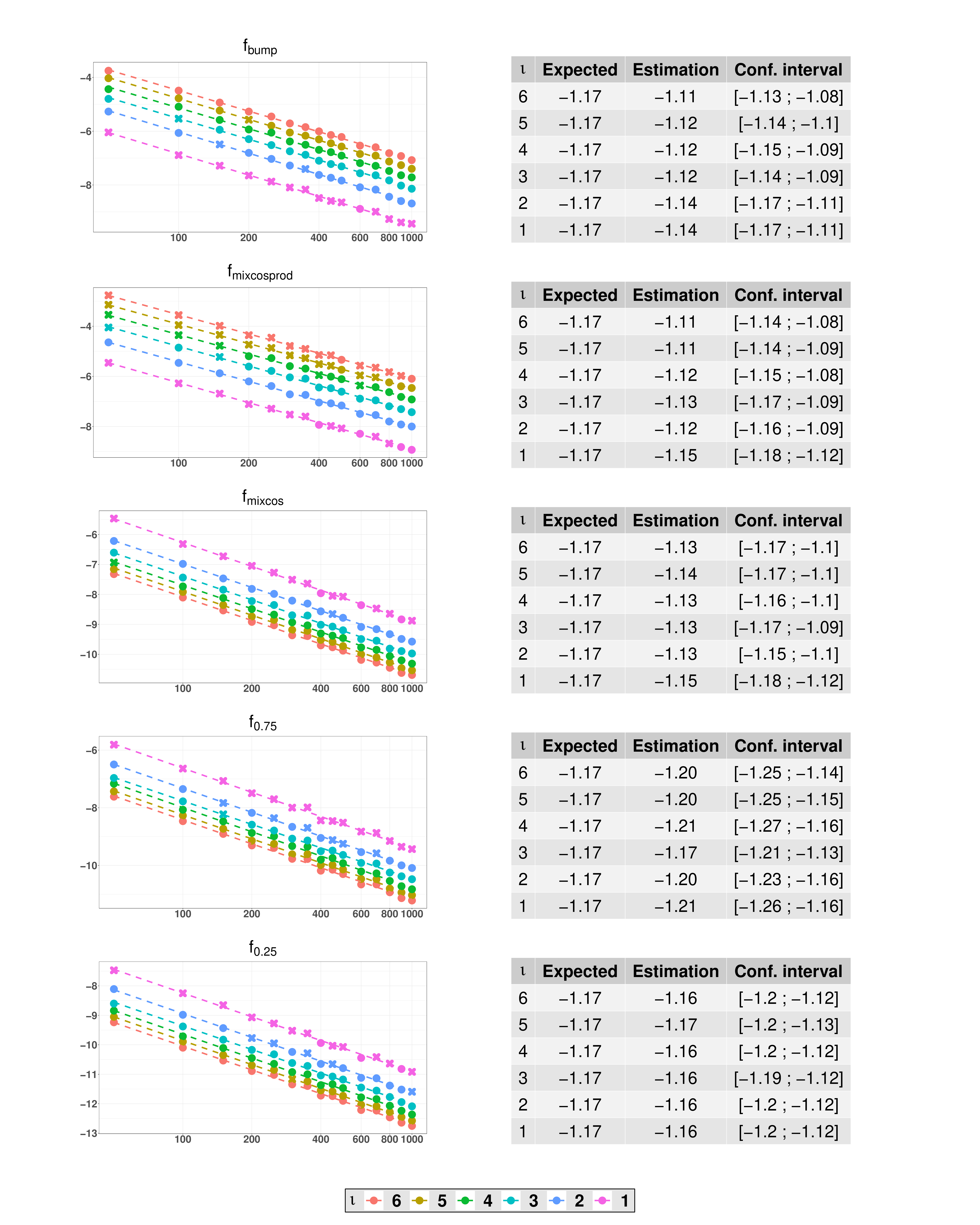}
	\caption{Summary of experiments in which integrals of $\iota$-dimensional functions are estimated by projecting a single $(N,6)$-Dirichlet DPP ($\iota=1\dots 6$). A $\bullet$ (resp. $\times$) indicates that the adjusted p-value of the Shapiro-Wilk test is not smaller (resp. smaller) than $5\%$.}
	\label{fig:ResMCCompWithProj6D}
\end{figure}

\begin{figure} 
	\vspace*{-2.5cm}
	\includegraphics[scale=.16]{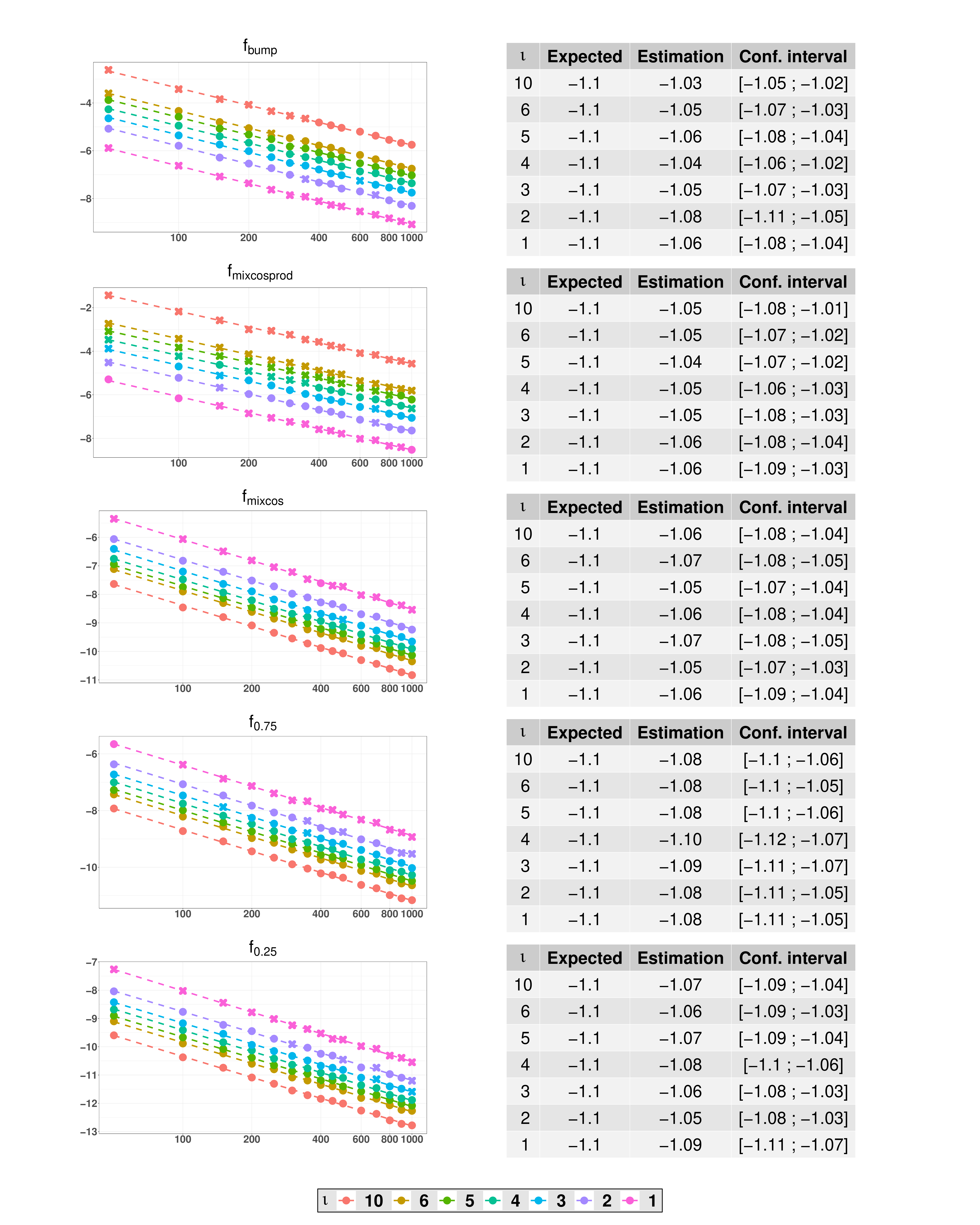}
	\caption{Summary of experiments in which integrals of $\iota$-dimensional functions are estimated by projecting a single $(N,10)$-Dirichlet DPP ($\iota=1\dots 6, 10$). A $\bullet$ (resp. $\times$) indicates that the adjusted p-value of the Shapiro-Wilk test is not smaller (resp. smaller) than $5\%$.}
	\label{fig:ResMCCompWithProj10D}
\end{figure}

We perform similar experiments. We set the dimension to $d=6$. For each configuration $\{x^{(1)},\dots,x^{(N)}\}$ of $(N,6)$-Dirichlet DPPs, we evaluate the estimator $\widehat \mu_N(f_I)$ given by~\eqref{eq:def_Iest} with $I\subset\overline{d}$ and $|I|=\iota=1,\dots,6$. In other words, we use $6$-dimensional configurations of points to estimate  integrals of $\iota$-dimensional integrands. 
Let us precise that a single realization $(N,6)$-Dirichlet is used for each value of $\iota$. However,
the  directions  kept when projecting are chosen randomly. 
Results are illustrated in Figure~\ref{fig:ResMCCompWithProj6D}, as for the previous case. 
The conclusions are quite similar to the previous case: the points remain nicely aligned along the regression lines and confidence intervals are in agreement with theoretical slopes,  all equal to $-1-1/6\approx -1.17$ in this situation. It seems that   normality is hardly hinted when a design is projected on low dimensional space: most of the values of $N$ are rejected by the Shapiro-Wilk tests for $\iota=1,2$. {We also conduct a similar experiment starting with realizations of the Dirichlet DPP in dimension $d=10$, see Figure~\ref{fig:ResMCCompWithProj10D}. Similar comments can be done from this situation.}

\subsection{Case $s<1/2$}

We now intend to illustrate Theorem~\ref{thm:asymp_Iest}~(i) for $d=1$ and $s<1/2$. We consider the following one-dimensional function defined on $\I{0}{1}$:
\begin{equation} \label{eq:ireg_func}
	h_\gamma(x) = \sum_{j\geqslant1}\frac{\cos(2\pi j (x-1/2))}{2\pi j^\gamma}.
\end{equation}
It is clear that $h_\gamma \in L^2_{\mathrm{per}}([0,1])$ if $\gamma>1/2$, and in such a case, $h_\gamma \in \mathcal H^s([0,1])$ for $s<\gamma-1/2$.

\begin{figure}
	\includegraphics[scale=.15]{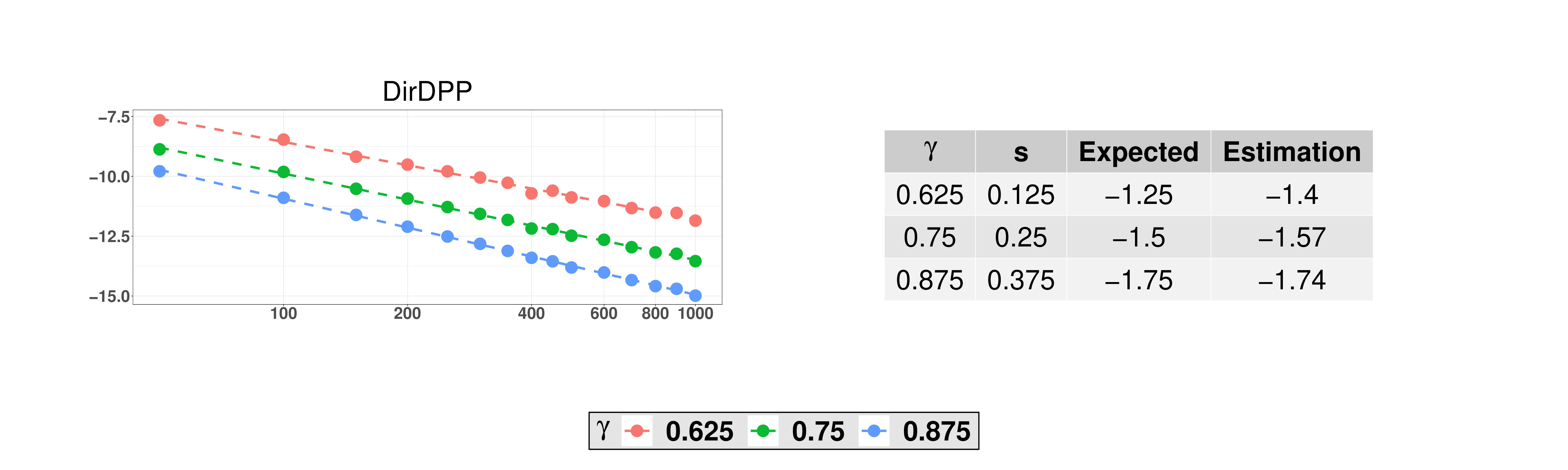}
	\caption{Summary of experiments in which integrals of function \eqref{eq:ireg_func} are estimated using $(N,1)$-Dirichlet DPP.}
	\label{fig:ResMCCompIreg}
\end{figure}

We repeat  experiments presented in Section~\ref{ssec:sim1}, but only for the case $d=1$: getting an accurate evaluation of~\eqref{eq:ireg_func} becomes really expensive for higher dimension. We consider three values for $\gamma$: 0.625, 0.75 and 0.875. Empirical results are depicted in Figure~\ref{fig:ResMCCompIreg}. As expected, the normality is rejected for all values of $N$ (for more readability we keep the dot representation for the points). Surprisingly, the results remains very satisfactory. When $\gamma=0.75,0.875$, the points remain well-aligned and the estimated slopes  are in good agreement with the theoretical $-1-2s/d$. The case $\gamma=0.625$ is less convincing but still very satisfactory, since $h_{0.625}$ is a very irregular function.

\section{A comparative numerical study.}
\label{sec:comparisons}
{We perform similar experiments with several designs: crude Monte-Carlo with uniform proposal, the DPP proposed by \cite{BardenetHardy20}  based on Legendre polynomials available in \texttt{DPPy} toolbox \cite{DPPYGautieretal19}, stratified sampling \cite{owen:book:13}, \textit{maximin} Latin Hypercube Design of \texttt{lhs} \texttt R package and Sobol and Halton sequences (\texttt{randtoolbox} \texttt R package). The Dirichlet DPP and these  different designs will respectively be denoted by \texttt{DirDPP}, \texttt{crude MC}, \texttt{BHDPP}, \texttt{stratified MC}, \texttt{maximinLHS}, \texttt{Sobol}, \texttt{Halton} in the following.}

\begin{figure} 
	\vspace*{-2.5cm}	\includegraphics[scale=.16]{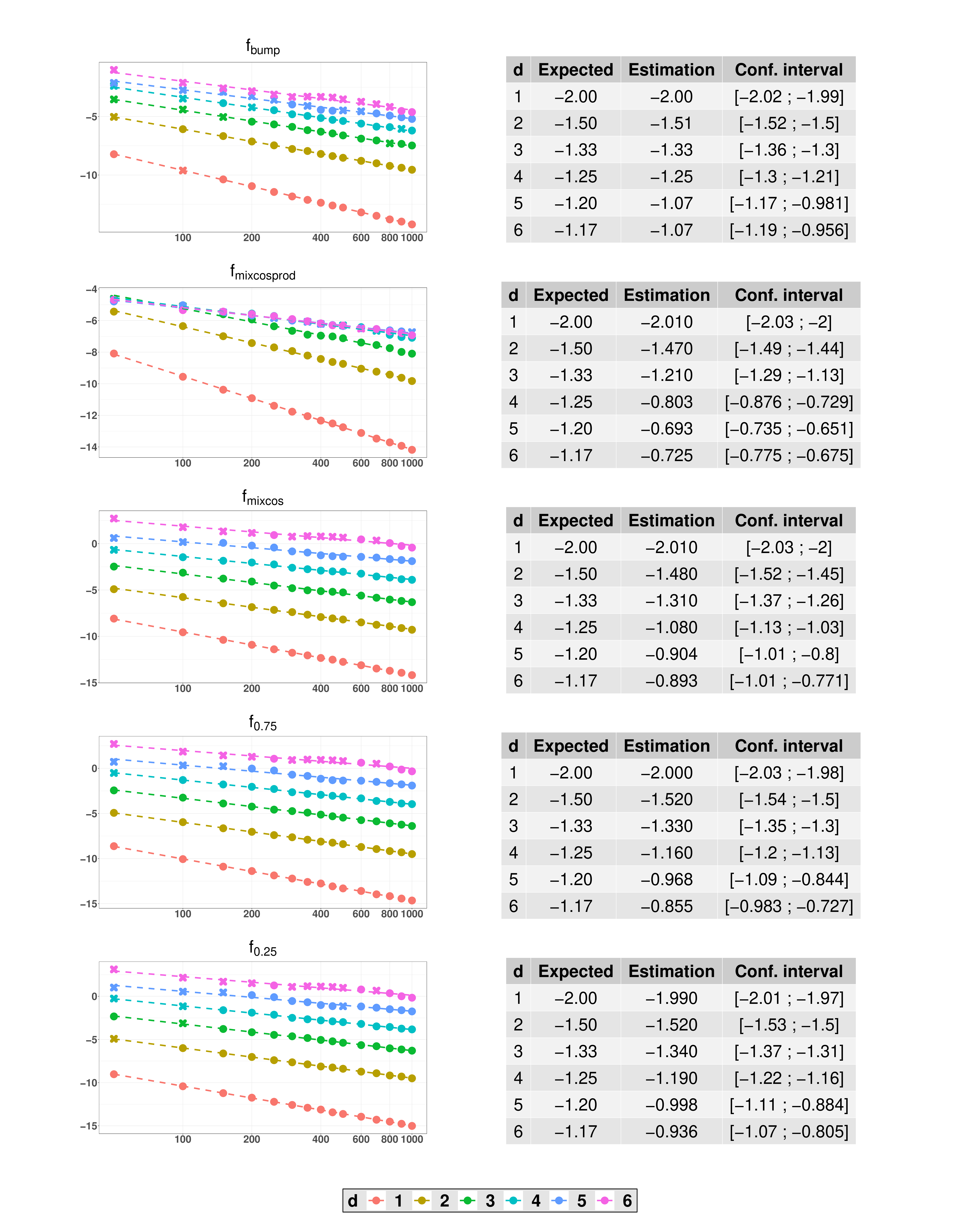}
	\caption{Summary of experiments in which integrals of $d$-dimensional functions are estimated using the \texttt{BHDPP} design.  A $\bullet$ (resp. $\times$) indicates that the adjusted p-value of the Shapiro-Wilk test is not smaller (resp. smaller) than $5\%$}
	\label{fig:ResMCCompBH}
\end{figure}

\begin{figure} 
	\vspace*{-2.5cm}\includegraphics[scale=.16]{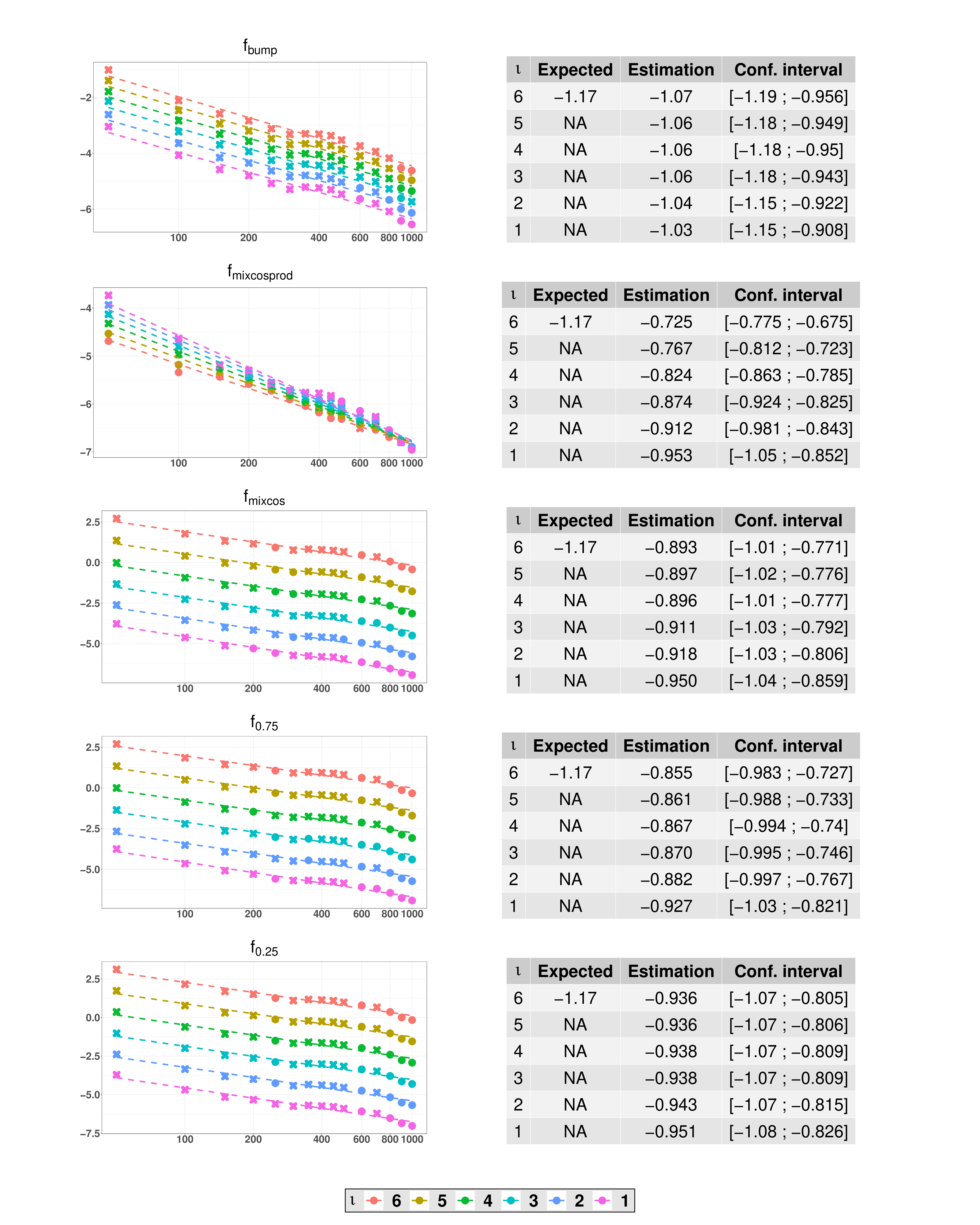}
	\caption{Summary of experiments in which integrals of $\iota$-dimensional functions are estimated by projecting a single 6-dimensionnal \texttt{BHDPP} design ($\iota=1\dots 6$). A $\bullet$ (resp. $\times$) indicates that the adjusted p-value of the Shapiro-Wilk test is not smaller (resp. smaller) than $5\%$.}
	\label{fig:ResMCCompWithProj6DBH}
\end{figure}

\begin{figure} 
	\includegraphics[scale=.15]{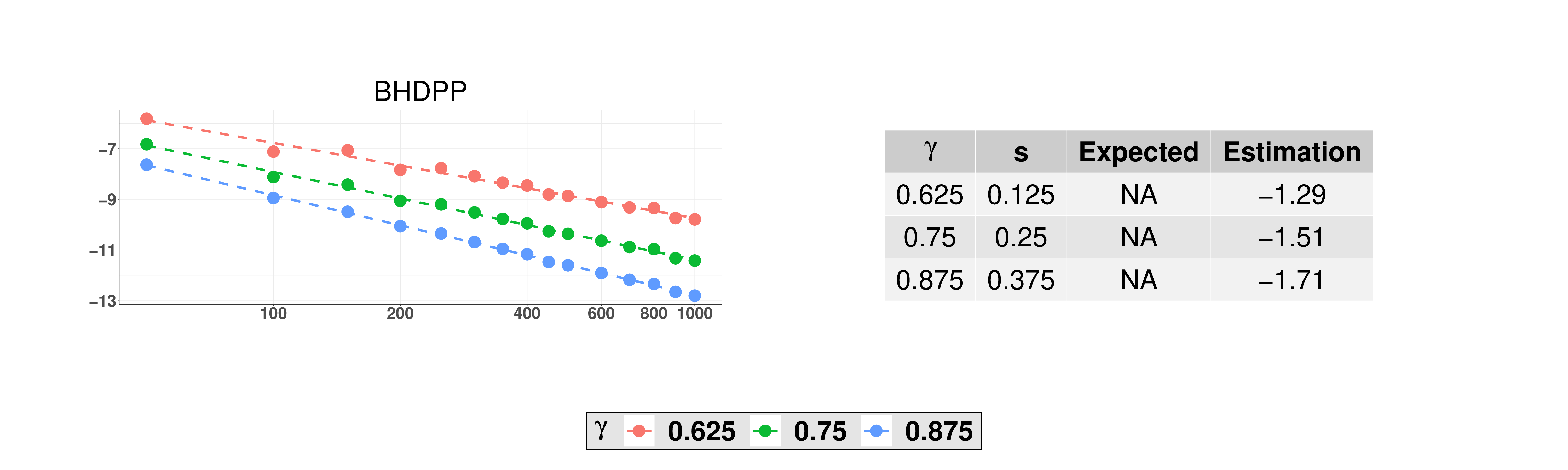}
 	\caption{Summary of experiments in which integrals of function \eqref{eq:ireg_func} are estimated using the BH design.}
 	\label{fig:ResMCCompIregBH}
 \end{figure}

{We first compare the \texttt{DirDPP} and \texttt{BHDPP} designs since they seem very close in terms of construction. Experiments presented in Section~\ref{sec:exp} (in the case $s>1/2$) for the \texttt{DirDPP} are conducted for the \texttt{BHDPP} design (only for $d=1,\dots,6$ to save time). Thus, Figures~\ref{fig:ResMCCompBH}-\ref{fig:ResMCCompIregBH} should be compared to  Figures~\ref{fig:ResMCComp},~\ref{fig:ResMCCompWithProj6D} and~\ref{fig:ResMCCompIreg}. Regarding Figures \ref{fig:ResMCComp} and \ref{fig:ResMCCompBH}, as expected, the results are quite similar for the $f_\text{bump}$ function, in terms of slope and confidence interval. However, we observe that for $d>4$, fluctuations around regression lines as well as confidence intervals are larger for the \texttt{BHDPP} design than for the \texttt{DirDPP} design. Moreover,  Gaussianity seems to be reached less often. 
Surprisingly, the \texttt{BHDPP} design gives satisfactory results even for non-differentiable integrands (including for function~\eqref{eq:ireg_func} according to Figure~\ref{fig:ResMCCompIregBH}), as long as $d<4$. But results deteriorate  for $d\ge 4$. The estimated slopes are even larger than $-1$ which corresponds to \texttt{crude MC} designs. \cite{BardenetHardy20} did not consider the problem of estimating $\iota$-dimensional integrals. Their point process was not designed to address such a problem, and this is indeed revealed by Figure~\ref{fig:ResMCCompWithProj6DBH}. When projections of a 6-dimensional \texttt{BHDPP} design is used to estimate $\iota$-dimensional integrals with $\iota=1,\dots,5$, the estimated slopes are at best close to -1 and at worse larger than -1. In all cases, the estimated slopes for the \texttt{BHDPP} design are far from $-1-1/6$ which on the contrary is reached by the \texttt{DirDPP} design.}

{Next, we compare the \texttt{DirDPP} design with more standard alternatives. For each design (\texttt{crude MC}, \texttt{strat MC}, \texttt{maximinLHS}, \texttt{Sobol}, \texttt{Halton}), we have reproduced exactly the same experiment which has led to Figures~\ref{fig:ResMCComp}-\ref{fig:ResMCCompWithProj10D}, except that due to the nature of the \texttt{maximinLHS}, \texttt{Sobol}, \texttt{Halton} designs, we have computed logarithms of the empirical mean squared errors (MSE) in terms of $\log(N)$. Of course, plotting the MSE  for the \texttt{DirDPP} design  instead of its variance  in  Figures~\ref{fig:ResMCComp}-\ref{fig:ResMCCompWithProj10D} would have led to the same conclusions, since estimates are unbiased (which turns out to be also the case for \texttt{crude MC}, \texttt{strat MC} and \texttt{BHDPP} designs). Moreover, similar figures summarizing empirical bias and variance have also been produced and are available on demand.}

\begin{figure}
\hspace*{-1.25cm}\includegraphics[width=1.2\textwidth]{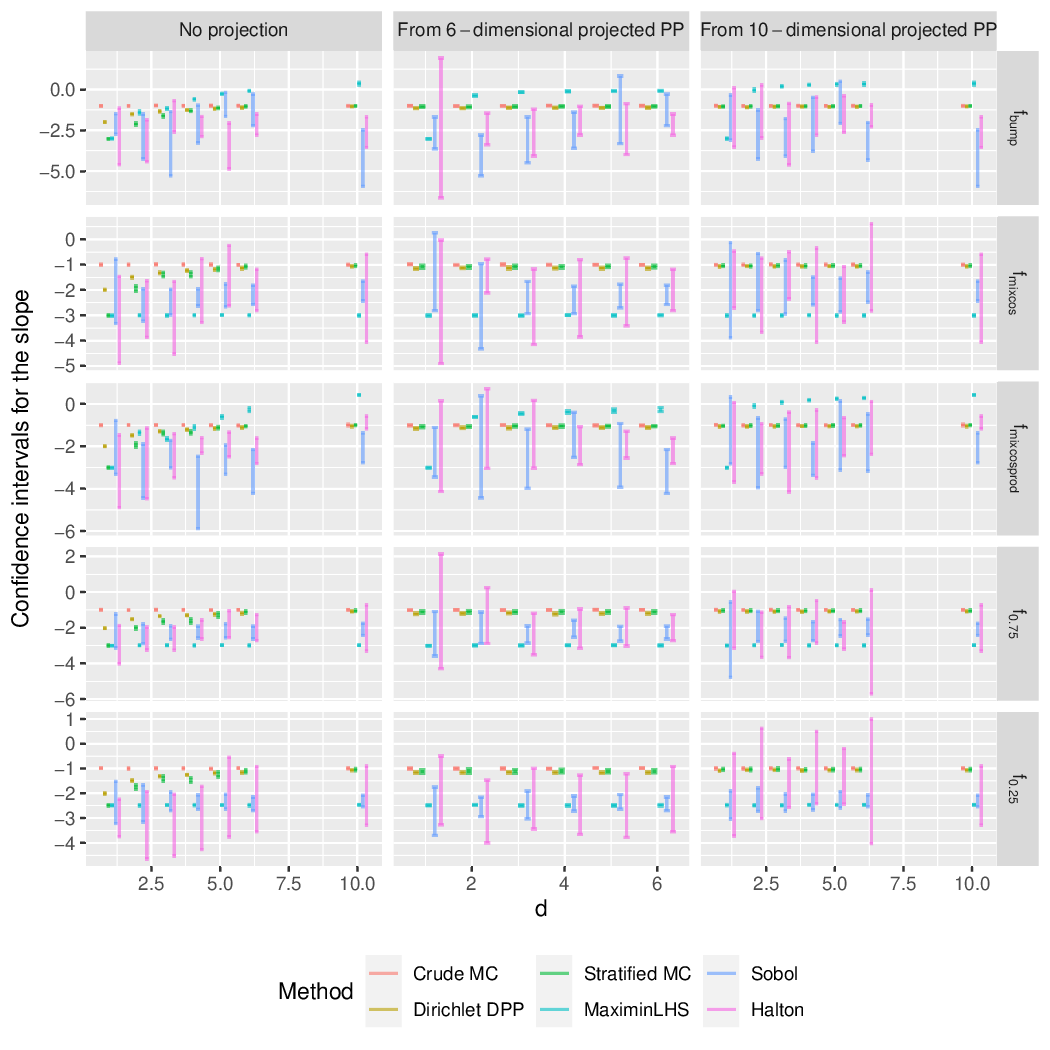}
\caption{Confidence intervals for the slope estimates of linear regressions of the logarithms of the MSE in terms of $\log(N)$ when we estimate $d$ dimensional integrals and designs generated in dimension $d$ (left) or $\iota$-dimensional integrals based on designs generated in dimension $d=6$ or $10$. Such experiments are done for the five test functions given by~\eqref{eq:fbump}-\eqref{eq:fgamma} and for different experimental designs.}
\label{fig:allSlopes}
\end{figure}

{Figure~\ref{fig:allSlopes} is 
a tentative summary of Figures~\ref{fig:ResMCComp}-\ref{fig:ResMCCompWithProj10D} and Figures~\ref{fig:crude1}-\ref{fig:halton3} presented in Appendix~\ref{app:simulations} for the sake of completeness. It reports confidence intervals for the slope estimates of linear regressions of the logarithms of the MSE in terms of $\log(N)$ for all designs and tests functions. 
All these figures are quite complex to comment. Our goal is not to point out which method is best for a particular function, or a particular sample size $N$, {\it etc}. Instead, we are willing to investigate if a method is ``stable'' (in terms of rate of convergence, or over a set of test functions, {\it etc}) in order to estimate $d$-dimensional integrals or $\iota$-dimensional integrals based on $d$-dimensional designs.}

{
\texttt{Crude MC} designs react as expected with a slope of -1 in all situations.} 

{\texttt{Stratified MC} designs seem to be quite competitive with respect to \texttt{DirDPP} designs, when one estimates $d$-dimensional integrals, even for non-differentiable integrands. Nevertheless, we observe that the rate of convergence is not exactly the same over the set of tests functions ({\it e.g}. $-3$ for the bump function and -2.5 for $f_{0.25}$ when $d=1$). Properties of \texttt{stratified MC} designs are significantly deteriorated when we estimate $\iota$-dimensional integrals (see Figures~\ref{fig:strat2} and~\ref{fig:strat3}). Estimates tend to converge but with a lower rate of convergence than estimates based on the \texttt{DirDPP} design.}

{\texttt{MaximinLHS} designs are quite impressive when $d=1,\dots,6,10$ for functions like $f_{\text{mixcos}}, f_{0.75}$ and $f_{0.25}$ but present some huge failures for functions $f_{\text{bump}}$ and $f_{\text{mixcosprod}}$. We have no explanation for the difficulties of this design to estimate integrals of functions of the form~\eqref{eq:f0prod}. Similar behaviours are observed when we estimate $\iota$-dimensional integrals (see Figures~\ref{fig:maximin2} and~\ref{fig:maximin3}). This apparent lack of convergence for the MSE is essentially explained by the fact that the empirical bias does not converge with $N$. }

{\texttt{Sobol} and \texttt{Halton} designs exhibit similar behaviours. Looking at Figures~\ref{fig:sobol1} and \ref{fig:halton1}, we observe that these designs lead to very small MSE (often much smaller than the ones obtained with other methods). However, what is striking when we examine Figure~\ref{fig:allSlopes} is the huge variability of slope estimates. Indeed, in Figures~\ref{fig:sobol1} and \ref{fig:halton1}, the (logarithms of) MSE have a high variance in particular for functions $f_{\text{mixcos}}, f_{0.75}$ and $f_{0.25}$ ({\it i.e.} for functions of the form~\eqref{eq:f0sum}). For these functions, results are difficult to explain as the dimension does not seem to affect the rate of convergence. Empirical results are even more variable when we are  interested in estimating $\iota$-dimensional integrals (see Figures~\ref{fig:sobol2}-\ref{fig:halton2} and Figures~\ref{fig:sobol3}-\ref{fig:halton3}). There is no clear rate of convergence over the set of test functions.}

{Overall, from this (clearly not exhaustive) empirical study, we do not claim that the \texttt{DirDPP} design outperforms other  designs investigated. For a specific sample size $N$, a specific function, a specific dimension, we may find a much better method. However, the \texttt{DirDPP} design is the only one (with the baseline \texttt{crude MC} design) which exhibits a clear and expected rate of convergence for any function, any dimension $d$ or any ``sub-dimension'' $\iota$.}

\section{Conclusion} \label{sec:conclusion}

In this paper, we build a specific class of repulsive point process and use its realization as  quadrature points to estimate integrals. The resulting Monte-Carlo estimator is unbiased  with a variance scaling as $\mathcal O (N^{-1-(2s \wedge 1)/d})$ when the integrand belongs to $\mathcal H^s([0,1]^d)$. Our methodology and results have the interest to be applied to non differentiable functions, an assumption which is often considered for other methods such as grid-based methods, RQMC, scrambled nets~\cite{owen:book:13}. We also show that the initial configuration of points can be used to estimate $\iota$-dimensional integrands ($\iota=1,\dots,d$) with the same efficiency than in dimension $d$. {To complete this work on a theoretical side, it could be interesting to investigate the rate of convergence to normality in some metric ({\it e.g. }Wassertstein or Kolmogorov-Smirnov). We have also left behind an important practical question: how to estimate $\sigma^2(f)$ or its approximation given by~\eqref{eq:est_asymp_const} and the matrix $\boldsymbol{\Sigma}_p$ given by~\eqref{eq:defSigmap}? Obviously, we do not want to assume known the Fourier expansion of $f$ nor to use external simulations to estimate such quantities which would make the current work definitely useless. Estimating $\sigma^2(f)$ or $\boldsymbol{\Sigma}_p$ using only the realization of the Dirichlet DPP appears to be highly challenging. Due to the intrinsic nature of this projection DPP, strategies like bootstrap, subsampling or leave-one out type procedures did not lead to a consistent estimate of the asymptotic variance. This problem is definitely an interesting perspective.}

We consider in this paper integrals on $[0,1]^d$ (or rectangles). Addressing the similar question for more complex domains or for improper integrals is of great interest. Similarly, considering integrands defined on manifolds such as torus or spheres is definitely an interesting perspective. {We could, for example, take advantage of the works by~\cite{moller:etal:2018} which presents several spherical-valued DPP models.}

{Crude Monte-Carlo or stratified Monte-Carlo designs can be improved by using antithetic or local antithetic methods, see \cite{owen:08,owen:book:13}. An idea of such improvements relies on the fact that $U\sim U(0,1)$ and $1-U$ have the same distribution. Such an  idea  could also be applied to our design:  This is again an advantage of the homogeneity of the Dirichlet DPP. We have not deeply studied these extensions but leave them for  future works. To give some flavor however, we can easily prove that the estimator
\[
		\widehat{\mu}_N^\prime( f_I) =  \frac1{2N} \sum_{j=1}^N \left\{ f_I ((u_j)_I) + f_I ((1-u_j)_I) \right\},
\]
where $1-u=(1-u_1,\dots,1-u_d)^\top$ for any $u \in [0,1]^d$, satisfies Theorem~\ref{thm:asymp_Iest} and Corollary~\ref{thm:asymp_IestProj}, with the same rate of convergence, the same assumptions on $f_I$.  However, the asymptotic constant is given by~\eqref{eq:sigmafOmega} where $|\hat f_I(j)|^2$ is replaced by $\mathrm{Re}(\hat f_I(j))^2$: This necessarily leads to a variance reduction. 
}

In this paper we did not exploit the form of the integrand $f$. ``Is it possible to improve the rate of convergence by designing an appropriate DPP which exploits the form of $f$?" seems to be a difficult but again interesting question. {We believe this is not feasible with the Dirichlet DPP proposed in this paper but the question remains open.}

{Another natural extension would be to consider a general target product measure $m$ on $[0,1]^d$, instead of the uniform distribution. The naive approach would be obviously to estimate $\int_{[0,1]^\iota} f_I m_I \de x$ with Dirichlet DPP. But, as \cite{BardenetHardy20} did, the measure $m$ could be used to build a dedicated DPP.
According to the proofs in our work, it seems reasonable to think that Theorem \ref{thm:asymp_Iest} and Corollary~\ref{thm:asymp_IestProj} could be extended to any product measure $m$ such that an orthonormal basis $\{\varphi_k(\cdot)\}_{k}$ of $L^2([0,1]^d,m)$ satisfies a separability property $\varphi_k(x)\varphi_l(x) = \varphi_{k+l}(x)$.} 

One of the limitation of the methodology lies in the simulation of a {continuous} DPP. Current algorithms have a computational cost $\mathcal O(N^3)$ and {are based on chain rules, where each point location is sampled with a rejection method, with evaluations of the acceptance ratio can be very costly.
 As the number of generated points increases, passing the rejection step becomes harder, and this step makes the sampler even more costly. A personal communication from \cite{lavancier:rubak:21} shows that in dimension one this rejection step can be significantly improved with a proposal adapted to the Fourier kernel. This result lets us hope that this proposal could be extended to the $d$-dimensional Dirichlet DPP.} Finally, it is worth reminding the reader that we use homogeneous point patterns which can be used to estimate any integral. Therefore the 2500 replications from {the Dirichlet DPP}, for $N=50,100,\dots,1000$ and $d=1,\dots,6$ can be used to evaluate any integral and are freely available upon request.

\section*{Acknowledgements} {The authors would like to sincerely thank the reviewers for the great interest they have shown in our work, for the  important number of suggestions and comments they made  which have significantly improved a previous version of the manuscript. The authors would also like to thank Guillaume Gautier and R\'emi Bardenet for the fruitful discussions and for sharing their code simulating the DPP defined in \cite{BardenetHardy20}. JF Coeurjolly and A Mazoyer were supported by National Research Council of Canada for this research.}

\printbibliography

\appendix


\section{Proofs}

\subsection{Proof of Proposition~\ref{prop:reductionvariance}}
\begin{proof}
Since $\XX$ is a projection kernel, in~\eqref{eq:Iest_DPP}, $\lambda_j = 0$ or $1$, $\rho_\YY=N$. Moreover, the trick is that the Fourier basis satisfies $\phi_j(u) \overline{\phi_k}(u) = \phi_{j-k}(u)$ for any $j,k \in \ZZ^d$ and $u\in B$. These facts reduce~\eqref{eq:Iest_DPP} to
\begin{equation}
	\VV\left[\widehat{\mu}_N(f)\right] = \frac1N \int_B f(u)^2 \de u 
	-\sum_{j,k \in E_N}  |\widehat f({j-k})|^2,
\end{equation}
whereby \eqref{eq:var1} is deduced using Parseval's identity.
Let us focus now on the second term of the right-hand side of~\eqref{eq:var1}.
When $d=1$, it is quite well-known that
\begin{equation} \label{eq:j-k1}
\sum_{j,k=0}^{N-1} \abs{\widehat f(j-k)}^2 = \sum_{\abs{l}\leqslant N-1} (N-\abs{l}) \abs{\widehat f(l))}^2.\end{equation}
The result is easily deduced using the fact that $E_N$ is a rectangular subset of $\ZZ^d$.
\end{proof}

\subsection{Proof of Theorem~\ref{thm:asymp_Iest}}

\begin{proof}
Let $i=1,\dots, d$ and $j_i \in \ZZ$ with $|j_i|\le n_i-1$. Then,
\begin{align*}
		\prod_{i=1}^d \left(n_i-\abs{j_i}\right) &= N \prod_{i=1}^d \left(1-\frac{\abs{j_i}}{n_i}\right) \\ 
		&= N \sum_{k=0}^d (-1)^k \sigma_k\left(\frac{\abs{j_1}}{n_1},\dots,\frac{\abs{j_d}}{n_d}\right)	\\
		& = N\left(1-\sum_{i=1}^d\frac{\abs{j_i}}{n_i} + \sum_{1\leqslant i_1<i_2<d}\frac{\abs{j_{i_1} j_{i_2}}}{n_{i_1} n_{i_2}}+\dots+ (-1)^d\prod_{i=1}^d \frac{\abs{j_i}}{n_i}\right)
	\end{align*}
	where, for $k=1\dots d$, $\sigma_k(x_1,\dots,x_d)$ denotes the $k$-th elementary symmetric polynomials in $d$ variables:
\[
\sigma_k(x_1,\dots,x_d) = \sum_{1\leqslant i_1<\dots< i_k\leqslant d} x_{i_1}\dots x_{i_k}
\]
For $u,v\in \RR^d $ (with $u_\ell\neq 0, \,\ell=1,\dots,d$), we define $1/u$ (resp. $v/u$) by $1/u=(1/u_1,\dots,1/u_d)^\top$ (resp. $v/u=(v_1/u_1,\dots,v_d/u_d)^\top$). Similarly, $\|u\|_\infty$ stands for $\max_\ell |u_\ell|$.
It is clear that for any $k\ge 2$ and $j \in F_N =  \{j \in \ZZ^d: |j_i| \le n_i-1, \; i=1,\dots,d \}$
\[
\sigma_k \left(\frac{|j_1|}{n_1},\dots,\frac{|j_d|}{n_d} \right)  = 
\mathcal O\left(\left\|\frac{j}n\right\|_\infty^{2} \right) 
.\] 
Therefore, \eqref{eq:var2} can be rewritten as
\begin{align} 
\VV\left[\widehat{\mu}_N(f)\right]	 =& \frac{1}{N}\sum_{j\in F_N^c}\abs{\widehat f(j)}^2 +\frac{1}{N}\sum_{j\in F_N}\left\| \frac{j}n \right\|_1 \abs{\widehat f(j)}^2 \nonumber \\
&+  \frac1N \sum_{j \in F_N} R(j/n) \abs{\widehat f(j)}^2 \label{eq:var_prooftp}
\end{align}
where
\[
	R(j/n) = \left\{
\begin{array}{ll}
	0 & \text{ if } d=1\\
	\mathcal O\left(\left\|{\displaystyle \frac{j}n}\right\|_\infty^{2} \right) & \text{ if } d>1.
\end{array}
	\right.
\]
Under the asymptotic framework~\eqref{eq:cont_asymp}, we have that as $N\to \infty$
\begin{align}
\VV\left[\widehat{\mu}_N(f)\right]	 \sim& \frac{1}{N}\sum_{j\in F_N^c}\abs{\widehat f(j)}^2 +\frac{1}{N^{1+1/d}}\sum_{j\in F_N}\left\| \frac{j}\kappa \right\|_1 \abs{\widehat f(j)}^2 \nonumber \\
&+  \frac1{N^{1+2/d}} \sum_{j \in F_N} R(j/\kappa) \abs{\widehat f(j)}^2. \label{eq:varequivalent}
\end{align}
In the following, $\tau$ denotes a positive generic constant which may vary from line to line.

\noindent (i) Let $s\in (0,1/2)$ and $j \in F_N$,
\begin{align}
\left\| \frac j\kappa\right\|_1 & = \left\| \frac j\kappa\right\|_1^{2s} \left\| \frac j\kappa\right\|_1^{1-2s} \le \tau N^{\frac{1-2s}d} \left( 1+ \|j\|_\infty\right)^{2s}.
\end{align}
Hence,
\begin{align}
\frac1{N^{1+1/d}} \sum_{j\in F_N} \left\| \frac j\kappa\right\|_1 \abs{\widehat f(j)}^2 
&\le \frac{\tau}{N^{1+2s/d}} \sum_{j \in F_N} \left( 1+ \|j\|_\infty\right)^{2s} \abs{\widehat f(j)}^2 \nonumber\\
&=\mathcal O\left(N^{-1-2s/d}\right). \label{eq:i1}
\end{align}
Similarly,
\begin{align}
\frac1{N^{1+2/d}} \sum_{j \in F_N} R(j/\kappa) \abs{\widehat f(j)}^2 &\le \frac{\tau N^{(2-2s)/d}}{N^{1+2/d}} \sum_{j \in F_N} \left( 1+ \|j\|_\infty\right)^{2s} \abs{\widehat f(j)}^2	\nonumber\\
&=\mathcal O\left(N^{-1-2s/d}\right) \label{eq:i2}
\end{align}
and 
\begin{align}
\frac{1}{N}\sum_{j\in F_N^c}\abs{\widehat f(j)}^2 &\le
\frac{\tau}{N^{1+2s/d}}\sum_{j\in F_N^c} \left( 1+ \|j\|_\infty\right)^{2s}  \abs{\widehat f(j)}^2 \nonumber \\
& = o \left( N^{-1-2s/d} \right). \label{eq:i3}
\end{align}
Combining \eqref{eq:i1}-\eqref{eq:i3} with \eqref{eq:varequivalent} leads to the result.

\noindent (ii) We proceed similarly for the case $s>1/2$. Let us focus on the case $d>1$ as the other one is easily deduced. Since $f\in \mathcal H^s(B)$ for $s>1/2$, we have that as $N\to \infty$
\begin{align}
\frac1{N^{1+1/d}} \sum_{j\in F_N} \left\| \frac j\kappa\right\|_1 \abs{\widehat f(j)}^2 & \sim 	\frac1{N^{1+1/d}} \sum_{j\in \ZZ^d} \left\| \frac j\kappa\right\|_1 \abs{\widehat f(j)}^2. \label{eq:ii1}
\end{align}
We also have that for any $j\in F_N$, $R(j/\kappa) \le \tau (1+\|j\|_\infty)^{\min(2s,2) }$ which leads to
\begin{align}
	\frac1{N^{1+2/d}} \sum_{j \in F_N} R(j/\kappa) &\le \frac{\tau}{N^{1+2/d}} \sum_{j \in F_N} \left( 1+ \|j\|_\infty\right)^{\min(2s,2)} \abs{\widehat f(j)}^2	\nonumber\\
&=\mathcal O\left(N^{-1-2/d}\right). \label{eq:ii2}
\end{align}
Finally,
\begin{align}
\frac{1}{N}\sum_{j\in F_N^c}\abs{\widehat f(j)}^2 &\le
\frac{\tau}{N^{1+2s/d}}\sum_{j\in F_N^c} \left( 1+ \|j\|_\infty\right)^{2s}  \abs{\widehat f(j)}^2 \nonumber \\
& = \mathcal O\left(N^{-1-2s/d}\right)
=o \left( N^{-1-1/d} \right). \label{eq:ii3}
\end{align}
Combining \eqref{eq:ii1}-\eqref{eq:ii3} with \eqref{eq:varequivalent} leads again to the result.

\noindent (iii) We consider again the case $d>1$. To achieve this step, we apply \cite[Theorem~1]{soshnikov:02} to the sequence of random variables $S_N(f)=\sum_{j=1}^N f(u_j)$. First, from~\eqref{eq:var_asymp}
\begin{equation} \label{eq:sosh1}
	\VV(S_N(f)) \sim N^2 \VV(\widehat{\mu}_N(f)) \sim N^{1-1/d} \sigma^2(f) \to \infty
\end{equation}
as $N\to \infty$ if $d>1$. Second, 
\begin{equation}
	\label{eq:sosh2}
	\|f\|_\infty = \mathcal O(1) = o\left( N^{\tau (1-1/d)}\right)
 \end{equation}
 for any $\tau>0$. Third, $\EE(S_N(|f|))= \mathcal O(N)$ and for $\delta=(1+\delta^\prime)/(1-1/d)$ for some $\delta^\prime>0$, we have $\VV( S_N(f))^\delta = \mathcal O(N^{\delta(1-1/d)}) = \mathcal O(N^{1+\delta^\prime})$, which implies that
 \begin{equation}
 	\label{eq:sosh3}
 	 \EE(S_N(|f|))=\mathcal O(\VV( S_N(f))^\delta).
 \end{equation}
 Equations~\eqref{eq:sosh1}-\eqref{eq:sosh3} are the key-ingredients of~\cite[Theorem~1]{soshnikov:02}, which proves that
 \[
 	\frac{S_N(f)- N \mu(f)}{\sqrt{\VV(S_N(f))}}  \to {N}(0,1)
 \]
 as $N\to \infty$ and yields the result.
\end{proof}

\subsection{Proof of Proposition~\ref{prop:sigma2}}

{The proof of this result is a direct consequence of the expression of $\sigma^2(f),\sigma^2_{\mathrm{std}}(f)$ and $\sigma^2_{\mathrm{strat}}(f)$ and the following lemma.}

{\begin{lemma}\label{lemma:sigma2} ${ }$ \\
(i) Let $f$ be defined as~\eqref{eq:f0prod} then for any $j\in \ZZ^d$, $\hat f(j) = \prod_{k=1}^d \hat f_0(j_k)$, whereby we deduce that for any $s\ge 0$
\[
	|\hat f(j)|^2 = \prod_{k=1}^d |\hat f_0(j_k)|^2 \quad \text{ and } \quad
	\|f\|_{\mathcal H^s([0,1]^d)}^2 = d \, \|f_0\|_{\mathcal H^s([0,1])}^2 
	\|f_0\|_{\mathcal H^0([0,1])}^{2d-2}.
\]
(ii) Let $f$ be defined as~\eqref{eq:f0sum} then for any $j\in \ZZ^d$ 
\[
\hat f(j) = \sum_{k=1}^d \hat f_0(j_k)\delta_{0,j_{-k}}	 \quad \text{ where for } u,v\in \ZZ^m \quad	
\delta_{u,v} = \left\{
\begin{array}{ll}
1  & \text{if } u_1=v_1,\dots,u_m=v_m \\ 0 &\text{otherwise}	
\end{array}
\right.
\]	
and $j_{-k}=(j_1,\dots,j_{k-1},j_{k+1},\dots,j_d)$. Hence, for any $s\ge 0$
\[
	|\hat f(j)|^2 = \sum_{k=1}^d |\hat f_0(j_k)|^2 \delta_{0,j_{-k}} \quad \text{ and } \quad
	\|f\|^2_{\mathcal H^s([0,1]^d)} = d \|f_0\|^2_{\mathcal H^s([0,1])}.
\]
\end{lemma}}

{\begin{proof}
(i) The expressions for $\hat f(j)$ and $|\hat f(j)|^2$ are straightforward. From this we derive
\begin{align*}
\|f\|_{\mathcal H^s([0,1]^d)}^2&= 
\sum_{j_1\in \ZZ}\dots \sum_{j_d \in \ZZ} 
\{ |j_1|^{2s}+\dots + |j_d|^{2s}\} \prod_{k=1}^d|\hat f_0(j_k)|^2 \\
&= d \sum_{j_1\in \ZZ} |j_1|^{2s}|\hat f_0(j_1)|^2 \sum_{k=2}^d \sum_{j_k\in \ZZ} |\hat f_0(j_k)|^2\\
&= d \, \|f_0\|^2_{\mathcal H^{s}([0,1])} \|f_0\|^{2(d-1)}_{L^2([0,1])}.
\end{align*}	
(ii) Omitted.
\end{proof}}

\subsection{Proof of Corollary~\ref{thm:asymp_IestProj}}

\begin{proof}
Let $f_I^\uparrow:\RR^d\rightarrow\RR$ be the $d$-dimensional measurable function given by
$
f_I^\uparrow(x) = f_I(x_I)\ind{x_{I^c}\in B_{I^c}}.
$
Then 
\[
\widehat{\mu}_N(f_I^\uparrow) = \frac{1}{N} \sum_{u\in\XX}h(u) = \frac{1}{N} \sum_{u\in\XX_I} f_I(u) = \widehat{\mu}_N(f_I).
\]
Since $f_I\in\mathcal{H}^s(B_I)$, it is straightforward to see that $f_I^\uparrow\in \mathcal{H}^s(B)$. In particular
\begin{align*}
	\sum_{j\in\ZZ^d}(1+\norm{j}_\infty)^{2s}\abs{\widehat f_I^\uparrow(j)}^2 & = \sum_{j\in\ZZ^d}(1+\norm{j}_\infty)^{2s}\ind{j_{I^c}=\mathbf{0}_{d-\iota}}\abs{\widehat f_I({j_I})}^2 \\[2ex]
	& = \sum_{j\in\ZZ^\iota}^d(1+\norm{j}_\infty)^{2s}\abs{\widehat f_I(j)}^2 
\end{align*}
where for some $p\ge 1$, $\mathbf{0}_{p}$ is the zero vector in $\RR^p$.
Corollary~\ref{thm:asymp_IestProj} is deduced by applying Theorem~\ref{thm:asymp_Iest} to the function $f_I^\uparrow$.	
\end{proof}

\subsection{Proof of Corollary~\ref{cor:multivariate}}

\begin{proof}
Let us first note, as in the proof of Corollary~\ref{thm:asymp_IestProj} that $f_{I_\ell} \in \mathcal H^s(B_{I_\ell})$ is equivalent to say that $f_{I_\ell^\uparrow} \in \mathcal H^s(B)$. Moreover, following the proof of Theorem~\ref{thm:asymp_Iest}(i), it is shown that as $N\to \infty$
\[
	N^{1+1/d} \; \Cov \left( \widehat\mu_{N}(f_{I_\ell}^\uparrow), \widehat\mu_{N}(f_{I_{\ell^\prime}}^\uparrow)\right) \to \left(\boldsymbol\Sigma_p \right)_{\ell \ell^\prime}.
\]
Now, we follow Cramèr-Wold device: let $a \in \RR^p$, and let $Z_a = a^\top \left( \widehat{\mu}_{N,p} - \mu_p\right)$. Then, $Z_a = \widehat \mu_N(g_a) -\mu(g_a)$ with $g_a(u) = \sum_{\ell=1}^p a_\ell f_{I_\ell}^\uparrow(u)$. The result is therefore deduced since $g_a \in \mathcal H^s(B)$, $N^{1+1/d} \VV(\widehat \mu_N(g_a)) \to a^\top \boldsymbol \Sigma_p a$ and Theorem~\ref{thm:asymp_Iest}(ii) can be applied to $g_a$, that is as $N\to \infty$
\[
		\sqrt{N^{1+1/d}} \; a^\top \left( \widehat{\mu}_{N,p} - \mu_p\right) \to N(0, a^\top \boldsymbol \Sigma_p a).
\]	
in distribution.
\end{proof}


\section{Alternative proof of Theorem~\normalfont{\ref{thm:asymp_IestProj}}} \label{sec:appendix}
Here we propose an alternative proof of Corollary~\ref{thm:asymp_IestProj} (ii)-(iii) based on the characterization of the projected point process $\XX_I$.
\begin{proof}
(ii) From Proposition~\ref{prop:projDirichlet} and in particular the characterization of $(\alpha)$-DPPs as the union of independent particular DPPs (see \cite{Houghetal06}), we have that 
\begin{equation} \label{eq:decomposition}
\widehat \mu_N(f_I) = \frac{1}{N_{I^c}} \sum_{j=1}^{N_{I^c}} \widehat \mu_{N_I,j}(f_I)			
\end{equation}
where for $j=1,\dots,N_{I^c}$
\[
	\widehat \mu_{N_I,j}(f_I) = \frac1{N_I} \sum_{u \in \YY_j} f_I(v)
\]
and where $\YY_1,\dots,\YY_{N_I,j}$ are iid $(N_I,\iota)$-Dirichlet DPPs, that is $\widehat \mu_{N_I,j}$ is nothing else than an average of unbiased estimators of $\mu(f_I)$ based on an $(N_I,\iota)$-Dirichlet DPP for which Theorem~\ref{thm:asymp_Iest} can now be applied.

In particular, using Theorem~\ref{thm:asymp_Iest} (ii), we have
\begin{align*}
\VV(\widehat \mu_N(f_I)) = \frac1{N_{I^c}} \VV( \widehat \mu_{N_I,1}(f_I)) 
&\sim	\frac{1}{N_{I^c}} \, \frac1{(N_I)^{1+1/\iota}} \varsigma^2(f_I)
\end{align*}
as $N_I\to \infty$, with
\[
\varsigma^2(f_I) = \sum_{j\in\ZZ^\iota} \left(\sum_{i\in I}\frac{\abs{j_i}}{\gamma_i}\right) \abs{\widehat f_I(j)}^2
\]
where for any $i\in I$
\[
\gamma_i = \lim_{N_I\to\infty} n_iN_I^{-1/\iota}.
\]
Now, since
\begin{equation} \label{eq:rel_gam_kap}
\gamma_i \underset{N_I \to\infty}{\sim} \kappa_i N^{1/d} N_I^{-1/\iota}
\end{equation}

where $\kappa_i$ are given by~\eqref{eq:cont_asymp}, we deduce that
\[
\VV(\widehat \mu_N(f_I)) \sim \frac1{N^{1+1/d}} \sum_{j\in\ZZ^\iota} \sigma^2(f_I)
\]
as $N_I\to\infty$ and where $\sigma^2(f_I)$ is given by \eqref{eq:sigmafOmega}, which yields the result.

(iii) We can observe from~\eqref{eq:decomposition} that
\[
	\widehat \mu_N(f_I) - \mu (f_I) = \frac{1}{N_{I^c}}\sum_{j=1}^{N_{I^c}} \left(\widehat \mu_{N_I,j}(f_I) - \mu (f_I)\right)
\]
From Theorem~\ref{thm:asymp_Iest} (ii) and for any $j\in I$
\[
Z_{N_I,j} =\sqrt{N_I^{1+1/\iota}} \frac{\widehat \mu_{N_I,j}(f_I) - \mu (f_I)}{\varsigma(f_I)}\to {N}(0,1)
\]
in distribution.
Then we apply Lindeberg-Feller theorem {(see {\it e.g. }\cite{billingsley:2008})} to establish that as $N_{I^c}\to \infty$ 
\[
	\frac1{\sqrt{N_{I^c}}} \sum_{j=1}^{N_I^c} Z_{N_I,j} \to {N}(0,1)
\]
in distribution.
Therefore, in distribution as $N\to \infty$
\[
	\sqrt{\frac{N_I^{1+1/\iota}}{N_{I^c}}} \; \frac{\widehat \mu_N(f_I) - \mu (f_I)}{\varsigma(f_I)} \; \to \; {N}(0,1)
\]
whereby we deduce the result thanks to \eqref{eq:rel_gam_kap}.
\end{proof}

\section{Simulation results} \label{app:simulations}

\begin{figure} 
	\vspace*{-2.5cm} \includegraphics[scale=.15]{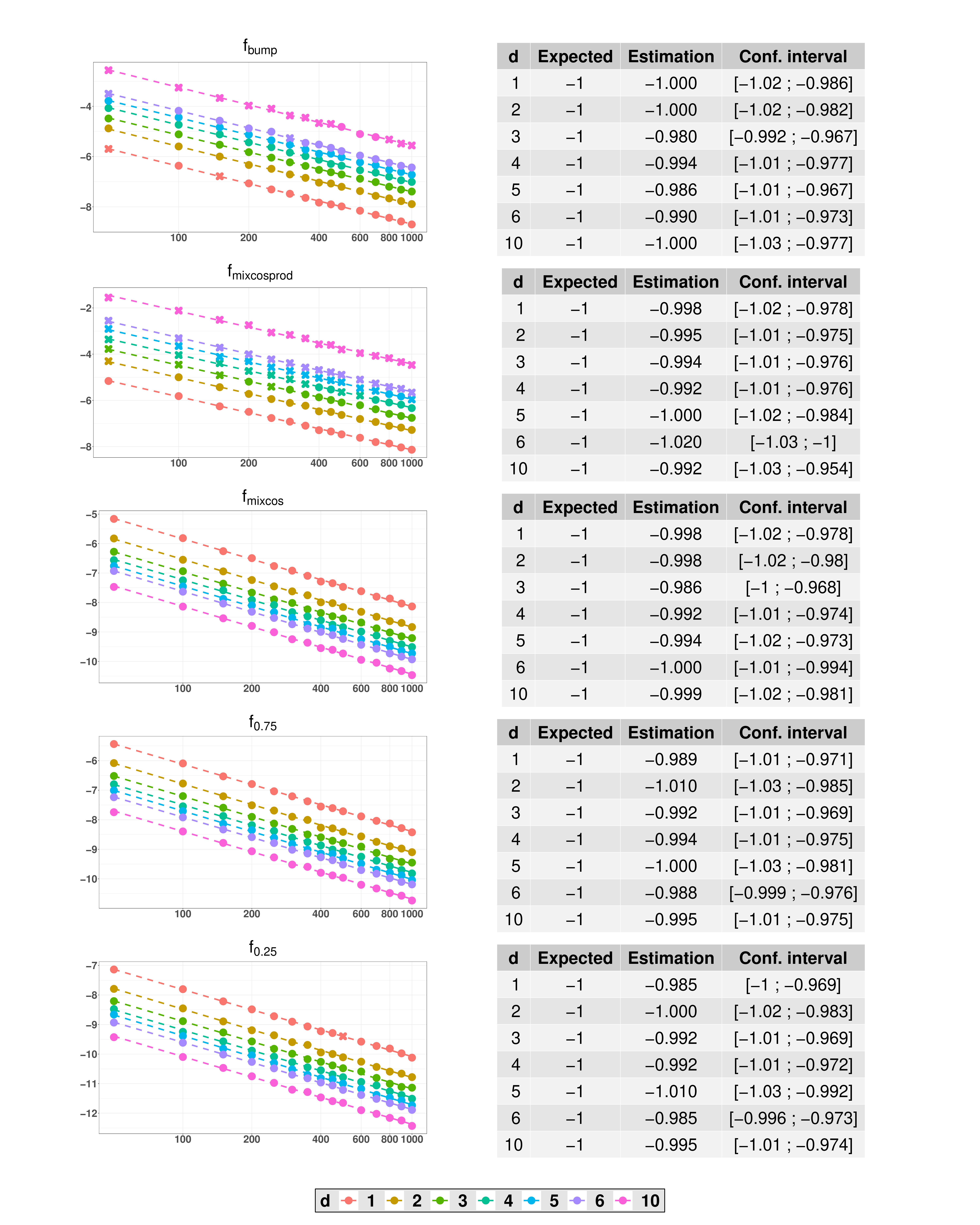}
	\caption{Summary of experiments in which integrals of $d$-dimensional functions are estimated using a \texttt{crude MC} design.  A $\bullet$ (resp. $\times$) indicates that the adjusted p-value of the Shapiro-Wilk test is not smaller (resp. smaller) than $5\%$}
	\label{fig:crude1}
\end{figure}

\begin{figure} 
	\vspace*{-2.5cm} \includegraphics[scale=.15]{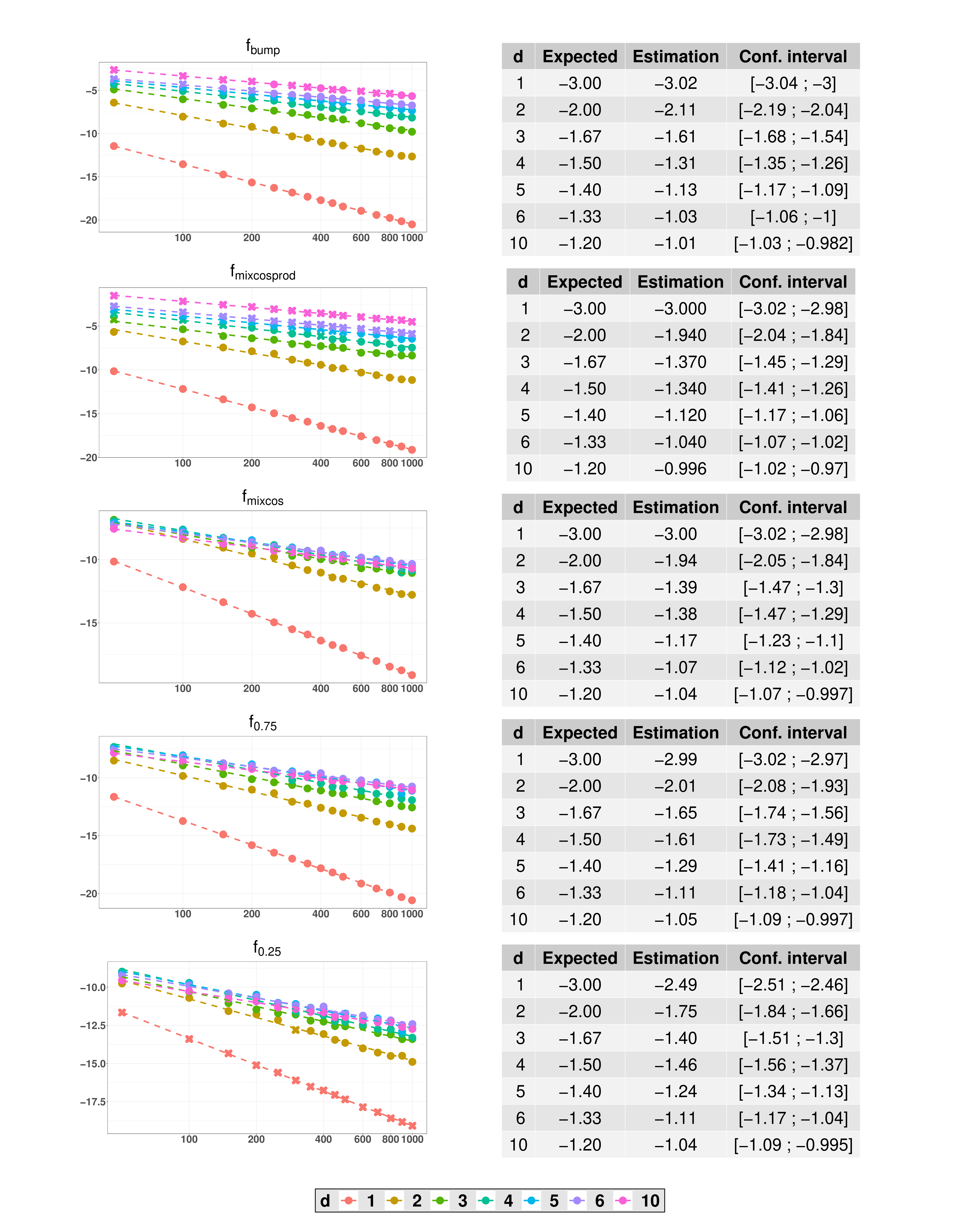}
	\caption{Summary of experiments in which integrals of $d$-dimensional functions are estimated using a \texttt{stratified MC} design.  A $\bullet$ (resp. $\times$) indicates that the adjusted p-value of the Shapiro-Wilk test is not smaller (resp. smaller) than $5\%$}
	\label{fig:strat1}
\end{figure}

\begin{figure} 
	\vspace*{-2.5cm}\includegraphics[scale=.15]{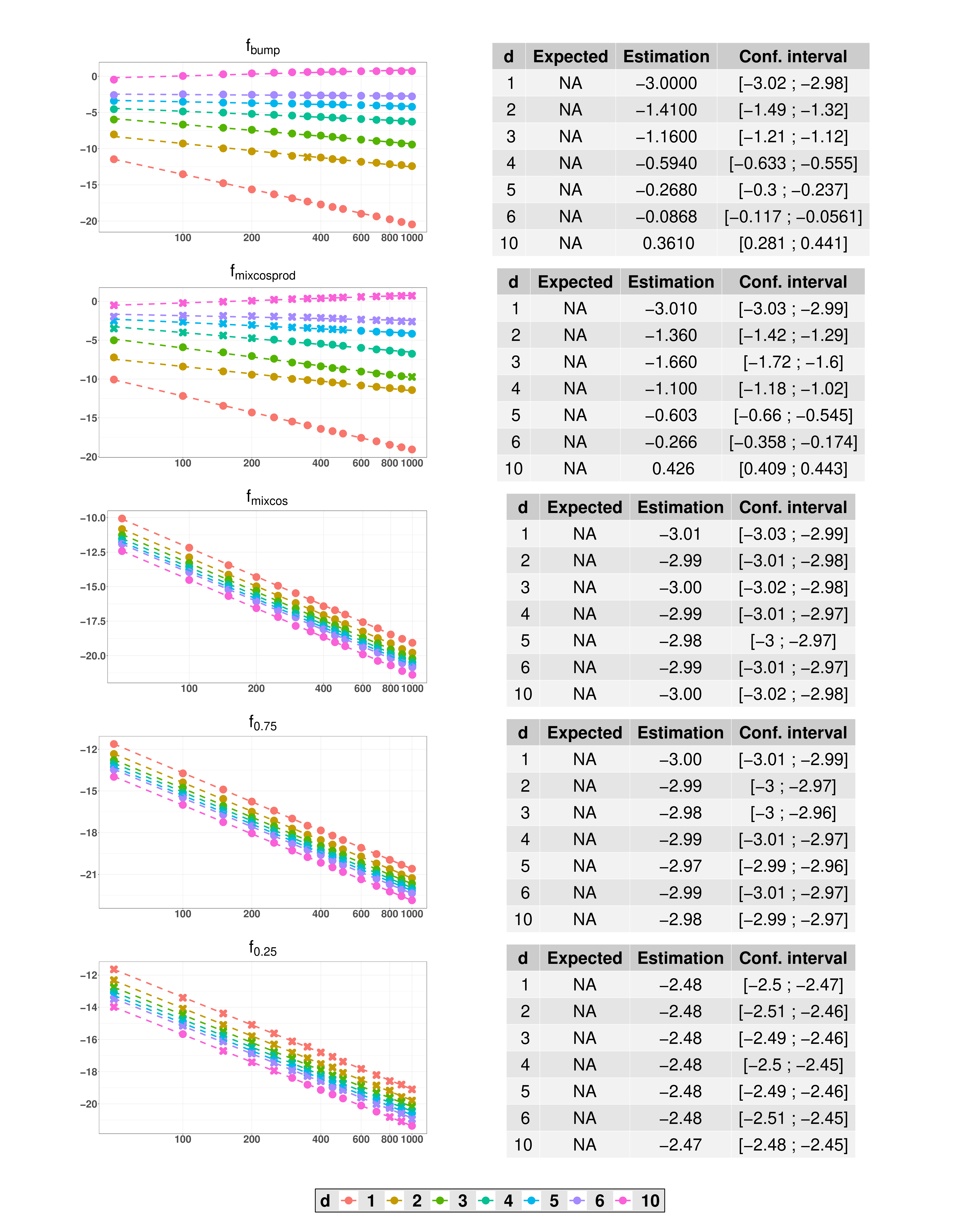}
	\caption{Summary of experiments in which integrals of $d$-dimensional functions are estimated using a \texttt{maximinLHS} design.  A $\bullet$ (resp. $\times$) indicates that the adjusted p-value of the Shapiro-Wilk test is not smaller (resp. smaller) than $5\%$}
	\label{fig:maximin1}
\end{figure}

\begin{figure} 
	\vspace*{-2.5cm}\includegraphics[scale=.15]{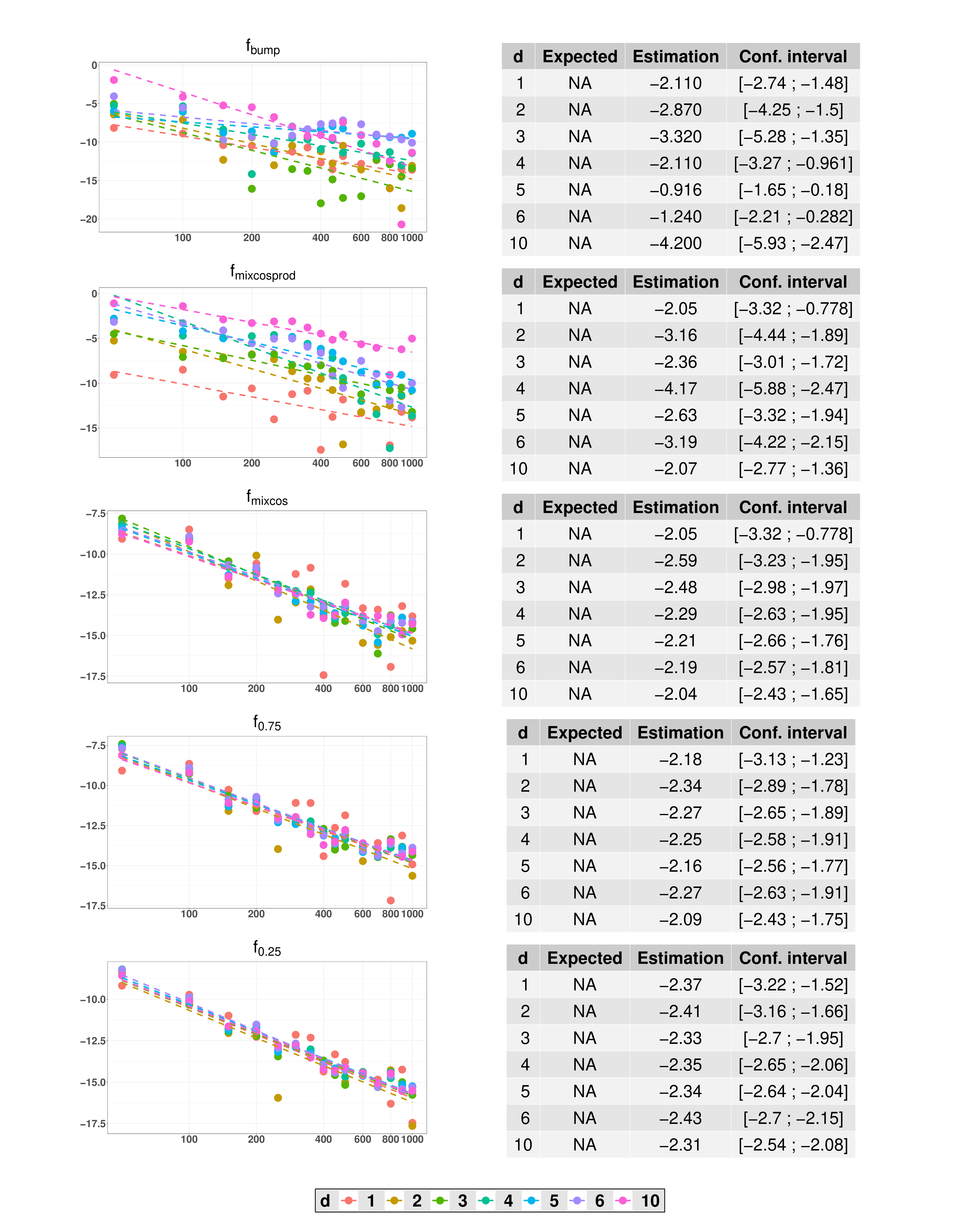}
	\caption{Summary of experiments in which integrals of $d$-dimensional functions are estimated using a \texttt{Sobol} design.}
	\label{fig:sobol1}
\end{figure}

\begin{figure} 
	\vspace*{-2.5cm}\includegraphics[scale=.15]{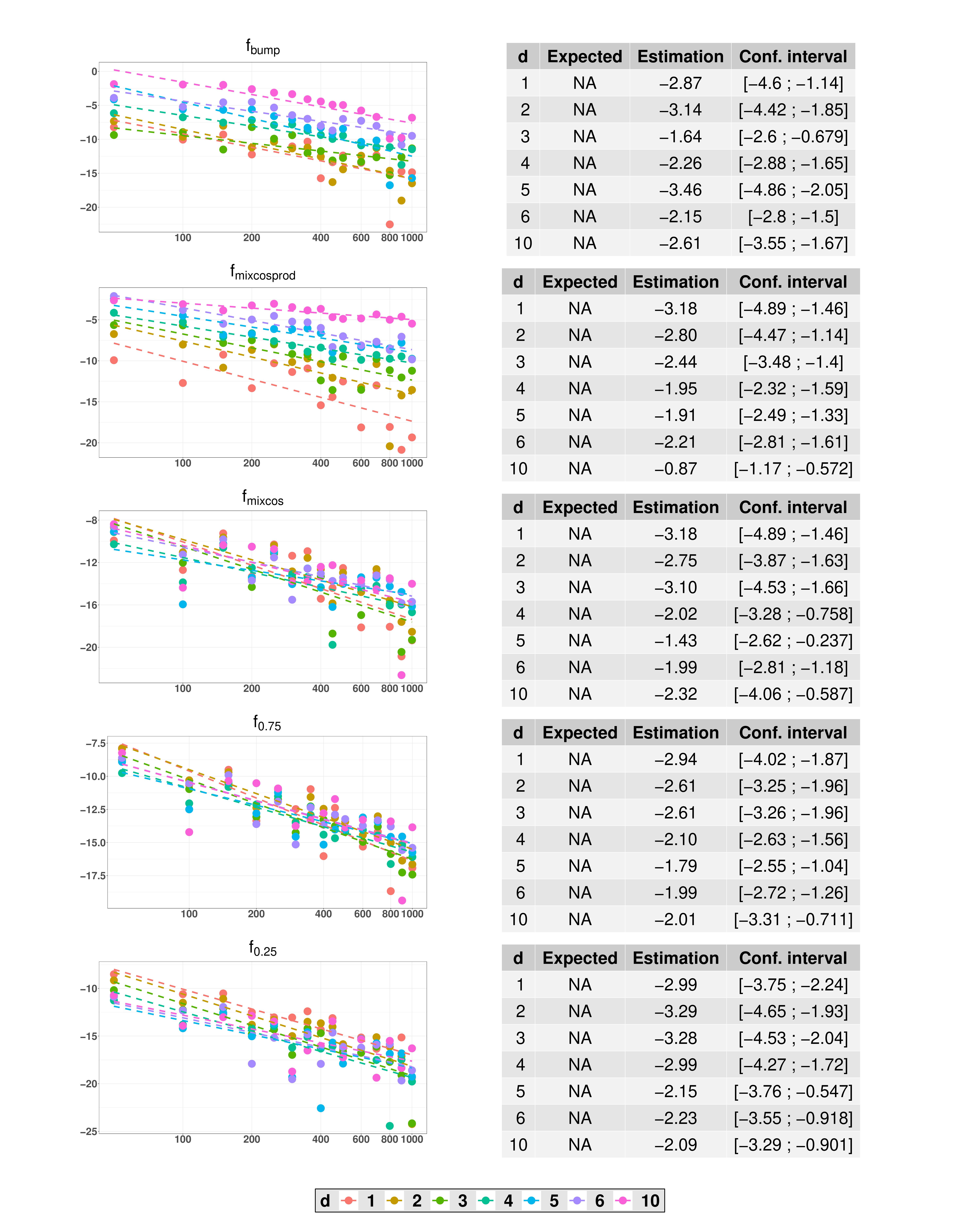}
	\caption{Summary of experiments in which integrals of $d$-dimensional functions are estimated using a \texttt{Halton} design.}
	\label{fig:halton1}
\end{figure}

\begin{figure} 
	\vspace*{-2.5cm}\includegraphics[scale=.16]{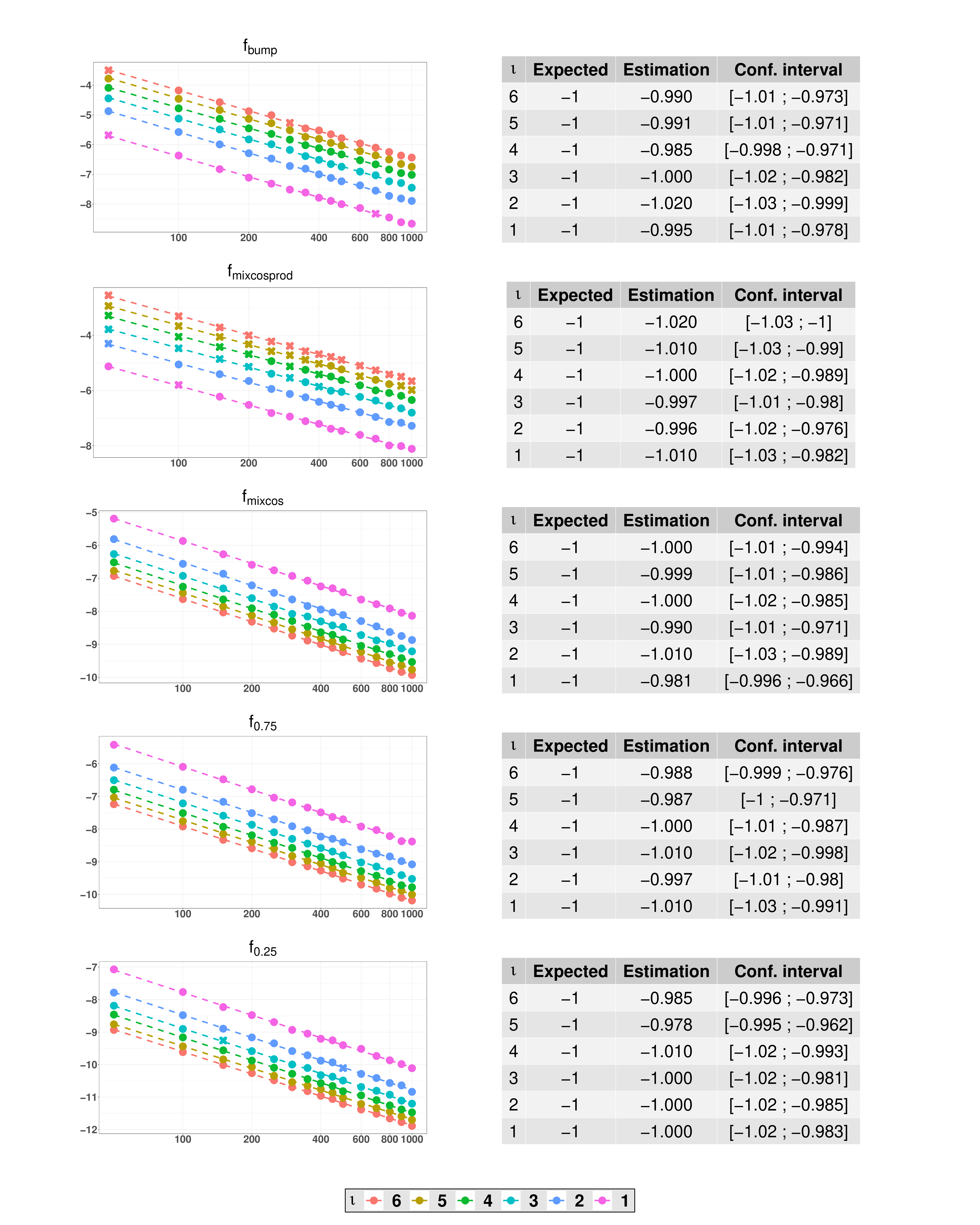}
	\caption{Summary of experiments in which integrals of $\iota$-dimensional functions are estimated by projecting a single 6-dimensional \texttt{crude MC} design ($\iota=1\dots 6$). A $\bullet$ (resp. $\times$) indicates that the adjusted p-value of the Shapiro-Wilk test is not smaller (resp. smaller) than $5\%$.}
	\label{fig:crude2}
\end{figure}

\begin{figure} 
	\vspace*{-2.5cm} \includegraphics[scale=.16]{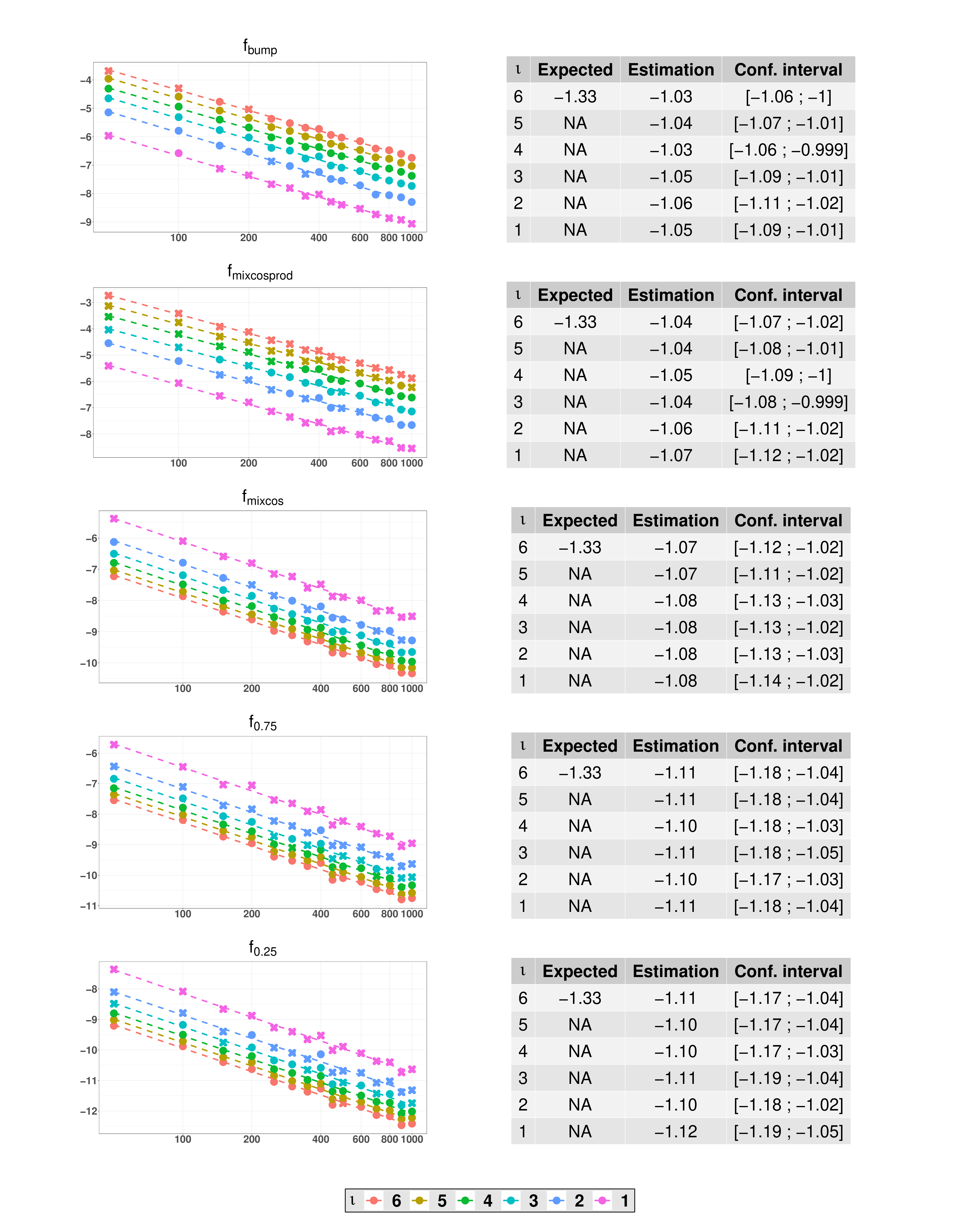}
	\caption{Summary of experiments in which integrals of $\iota$-dimensional functions are estimated by projecting a single 6-dimensional \texttt{stratified MC} design ($\iota=1\dots 6$). A $\bullet$ (resp. $\times$) indicates that the adjusted p-value of the Shapiro-Wilk test is not smaller (resp. smaller) than $5\%$.}
	\label{fig:strat2}
\end{figure}

\begin{figure} 
	\vspace*{-2.5cm}\includegraphics[scale=.16]{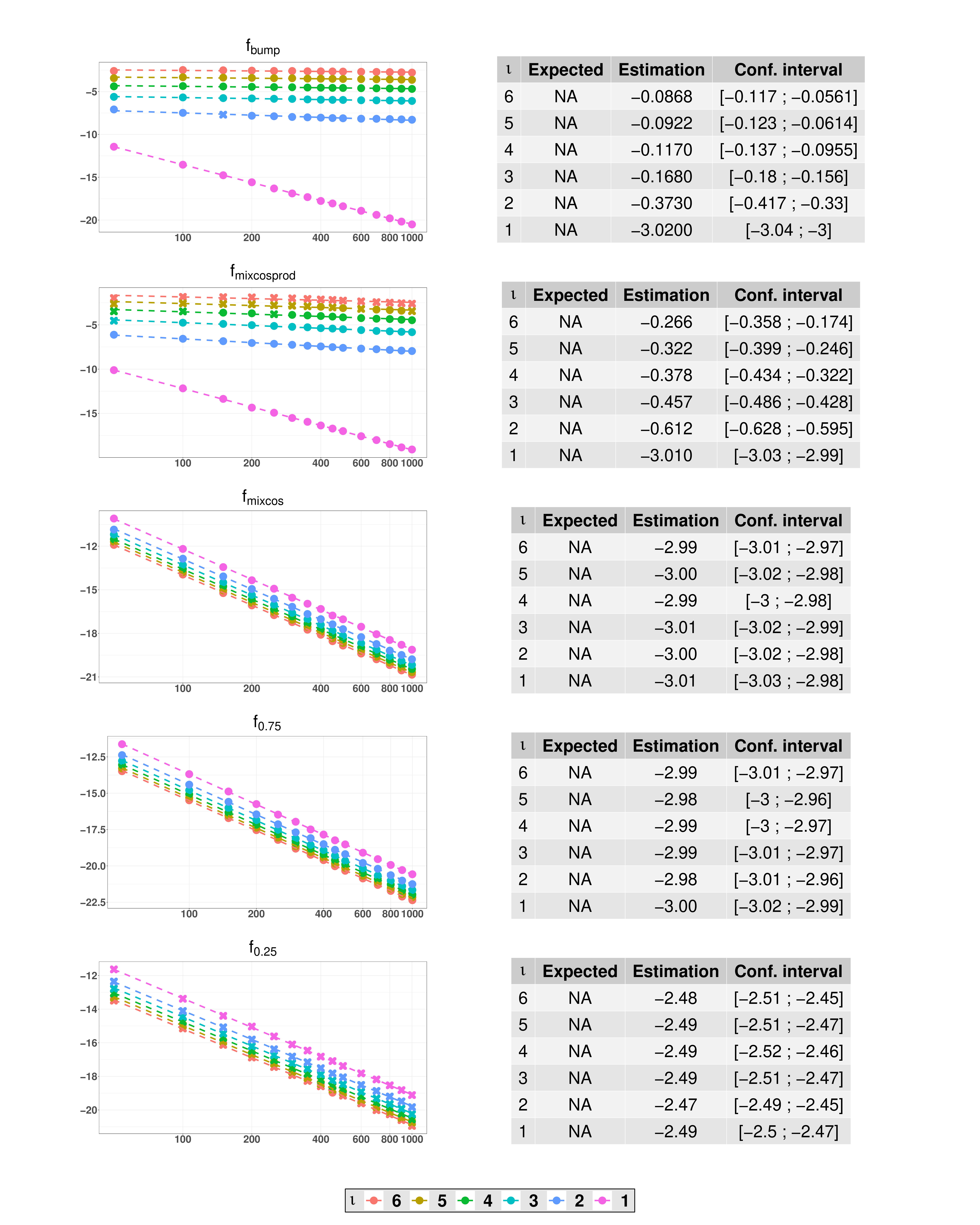}
	\caption{Summary of experiments in which integrals of $\iota$-dimensional functions are estimated by projecting a single 6-dimensional \texttt{maximinLHS} design ($\iota=1\dots 6$). A $\bullet$ (resp. $\times$) indicates that the adjusted p-value of the Shapiro-Wilk test is not smaller (resp. smaller) than $5\%$.}
	\label{fig:maximin2}
\end{figure}

\begin{figure} 
	\vspace*{-2.5cm}\includegraphics[scale=.16]{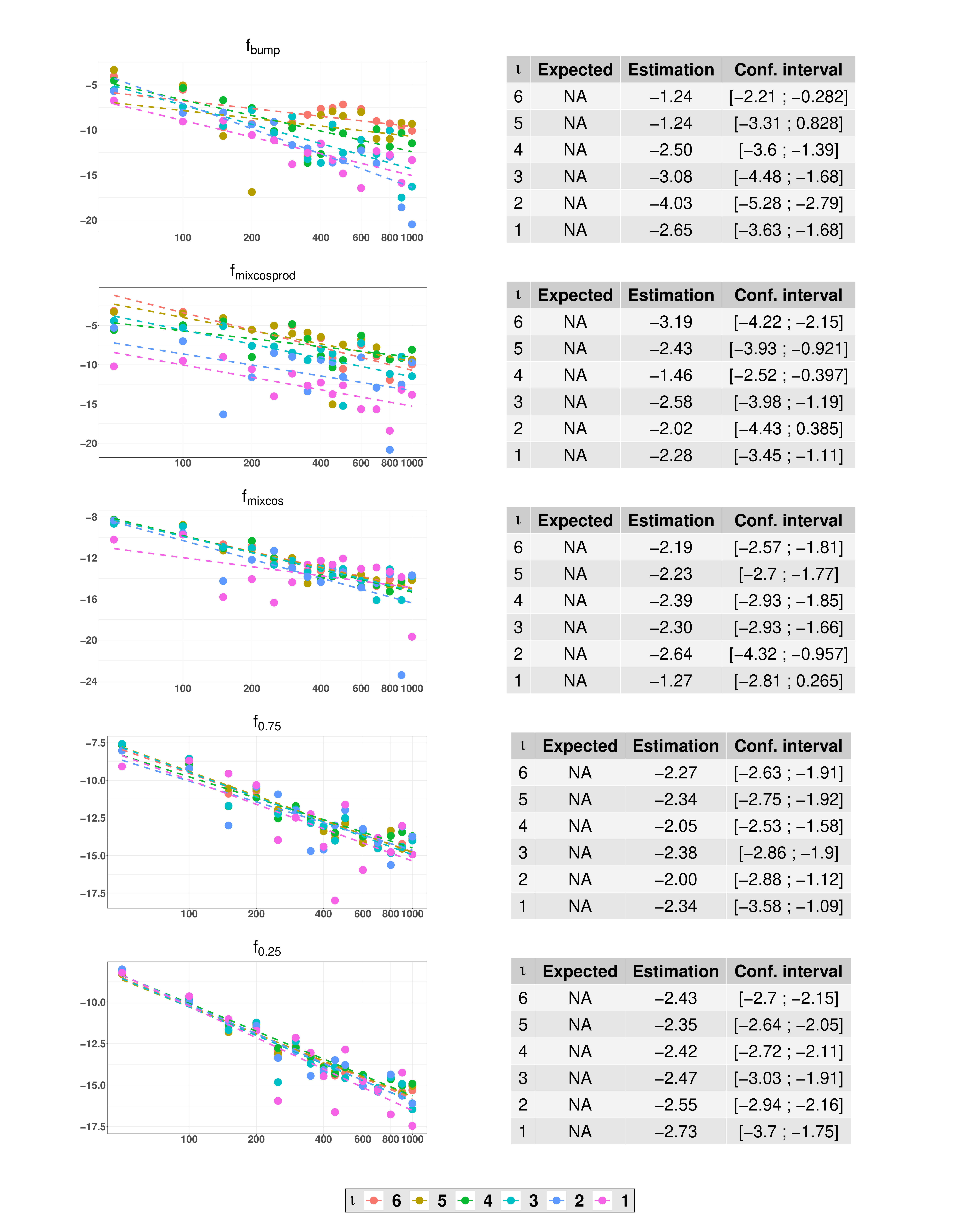}
	\caption{Summary of experiments in which integrals of $\iota$-dimensional functions are estimated by projecting a single 6-dimensional \texttt{Sobol} design ($\iota=1\dots 6$).}
	\label{fig:sobol2}
\end{figure}

\begin{figure} 
	\vspace*{-2.5cm}\includegraphics[scale=.16]{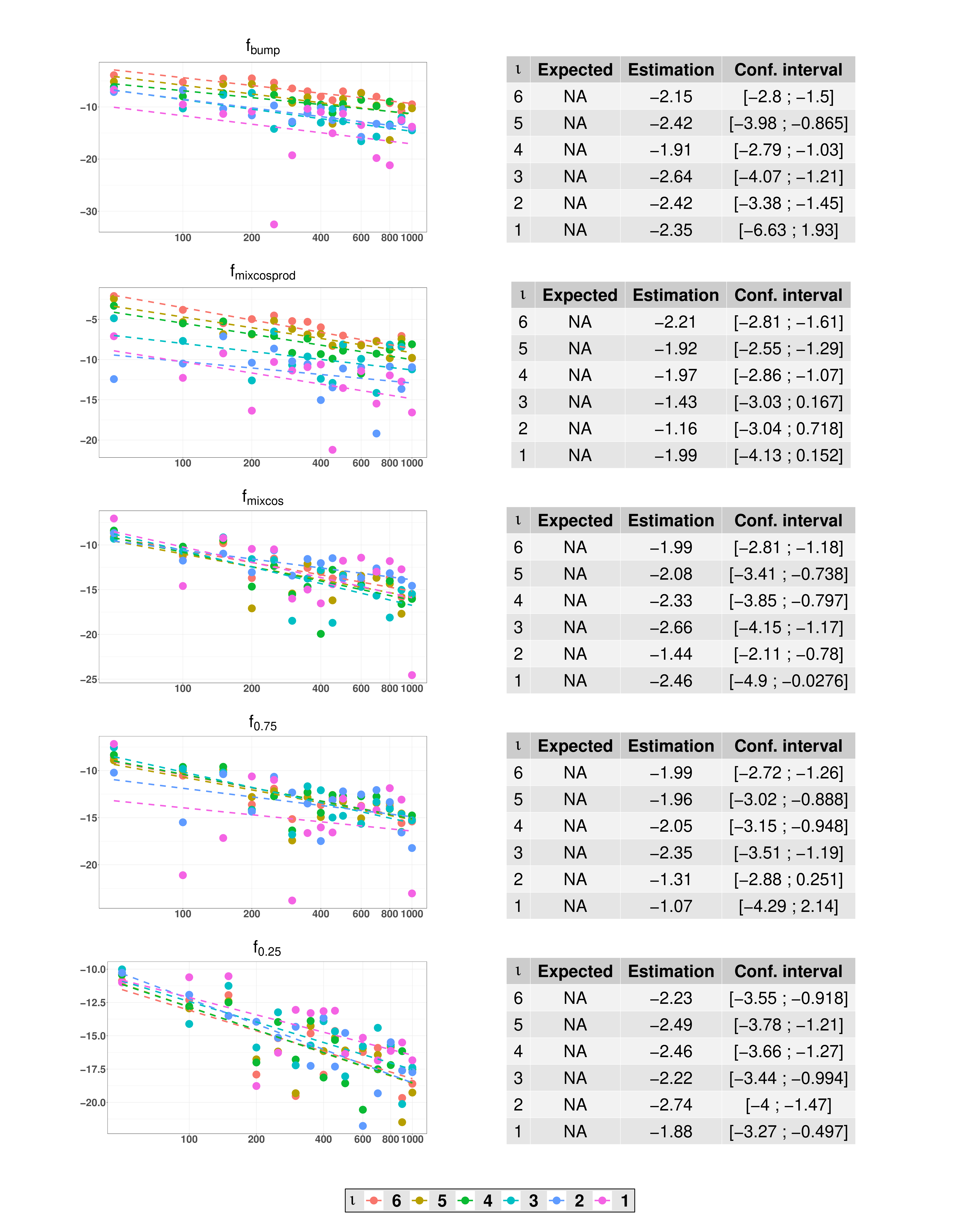}
	\caption{Summary of experiments in which integrals of $\iota$-dimensional functions are estimated by projecting a single 6-dimensional \texttt{Halton} design ($\iota=1\dots 6$).}
	\label{fig:halton2}
\end{figure}

\begin{figure} 
	\vspace*{-2.5cm}\includegraphics[scale=.16]{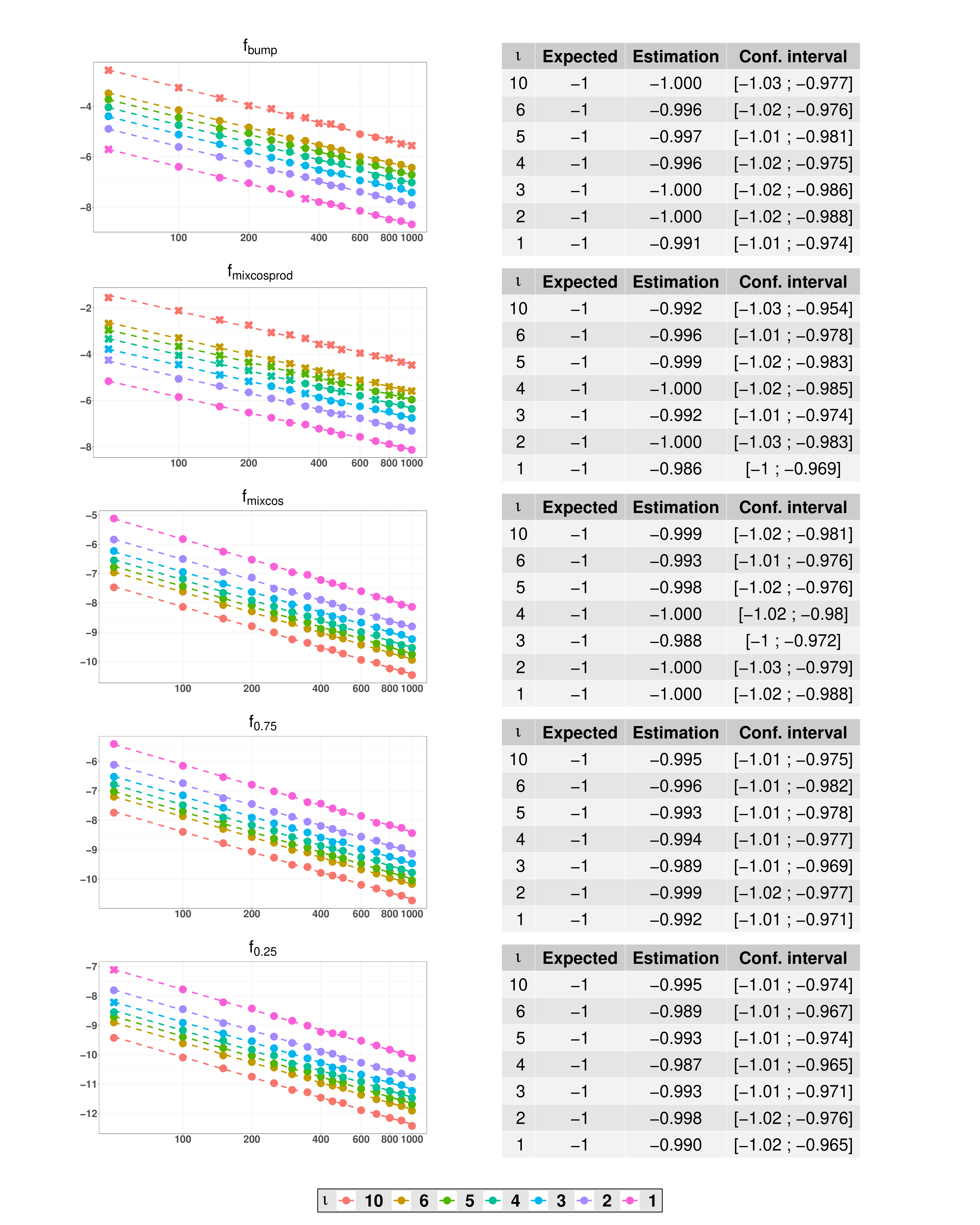}
	\caption{Summary of experiments in which integrals of $\iota$-dimensional functions are estimated by projecting a single 10-dimensional \texttt{crude MC} design ($\iota=1\dots 6,10$). A $\bullet$ (resp. $\times$) indicates that the adjusted p-value of the Shapiro-Wilk test is not smaller (resp. smaller) than $5\%$.}
	\label{fig:crude3}
\end{figure}

\begin{figure} 
	\vspace*{-2.5cm}\includegraphics[scale=.16]{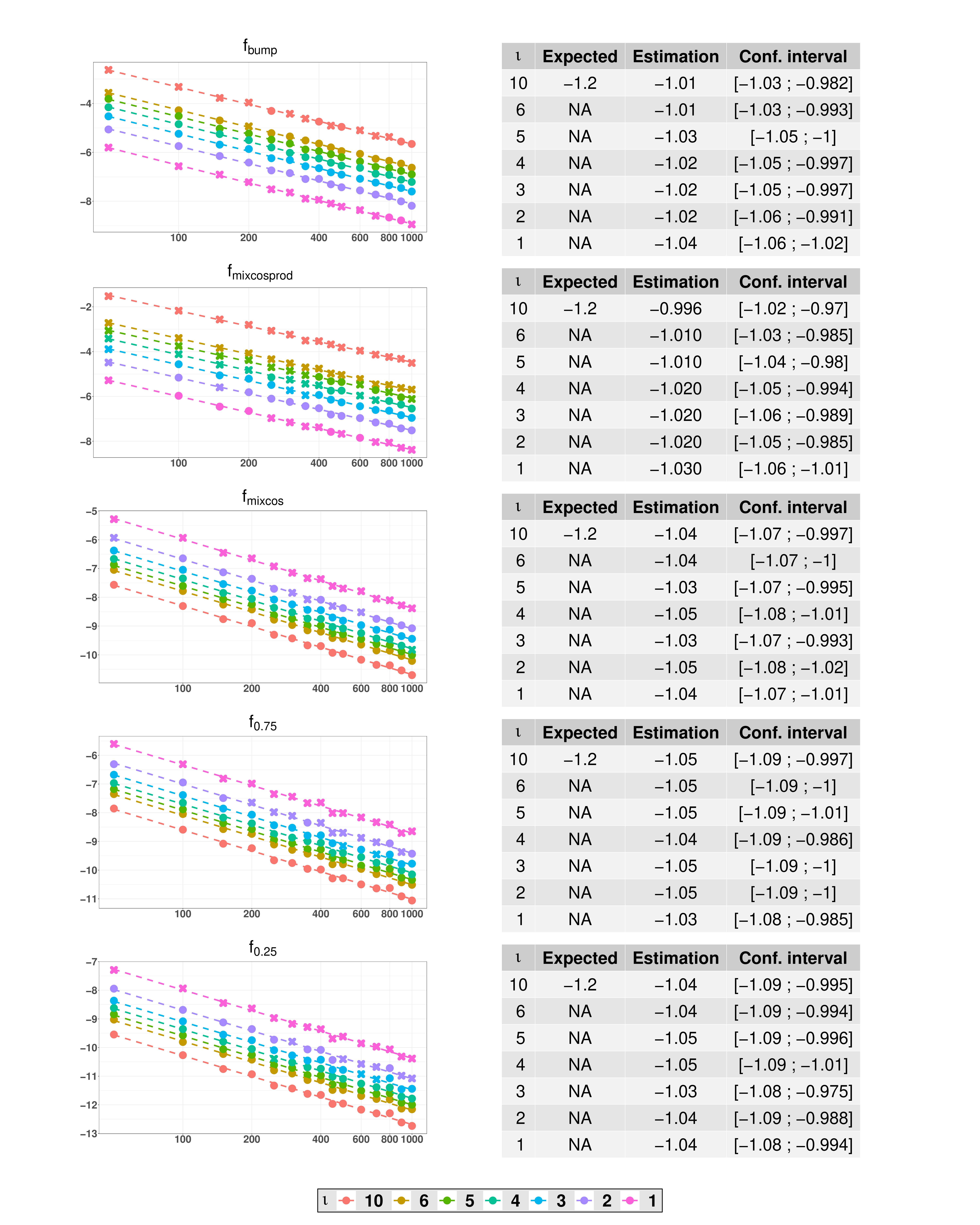}
	\caption{Summary of experiments in which integrals of $\iota$-dimensional functions are estimated by projecting a single 10-dimensional \texttt{stratified MC} design ($\iota=1\dots 6,10$). A $\bullet$ (resp. $\times$) indicates that the adjusted p-value of the Shapiro-Wilk test is not smaller (resp. smaller) than $5\%$.}
	\label{fig:strat3}
\end{figure}

\begin{figure} 
	\vspace*{-2.5cm}\includegraphics[scale=.16]{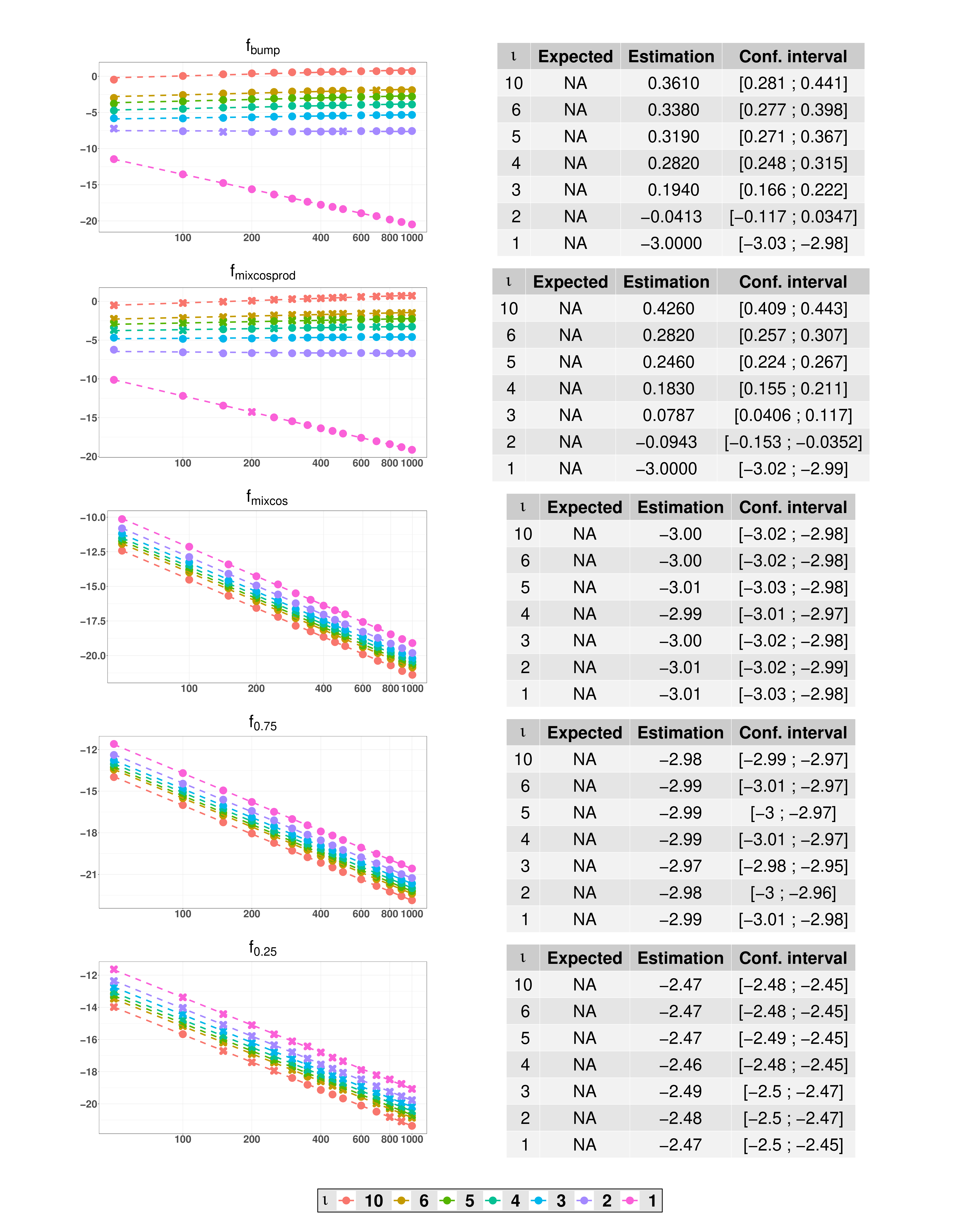}
	\caption{Summary of experiments in which integrals of $\iota$-dimensional functions are estimated by projecting a single 10-dimensional \texttt{maximinLHS} design ($\iota=1\dots 6,10$). A $\bullet$ (resp. $\times$) indicates that the adjusted p-value of the Shapiro-Wilk test is not smaller (resp. smaller) than $5\%$.}
	\label{fig:maximin3}
\end{figure}

\begin{figure} 
	\vspace*{-2.5cm}\includegraphics[scale=.16]{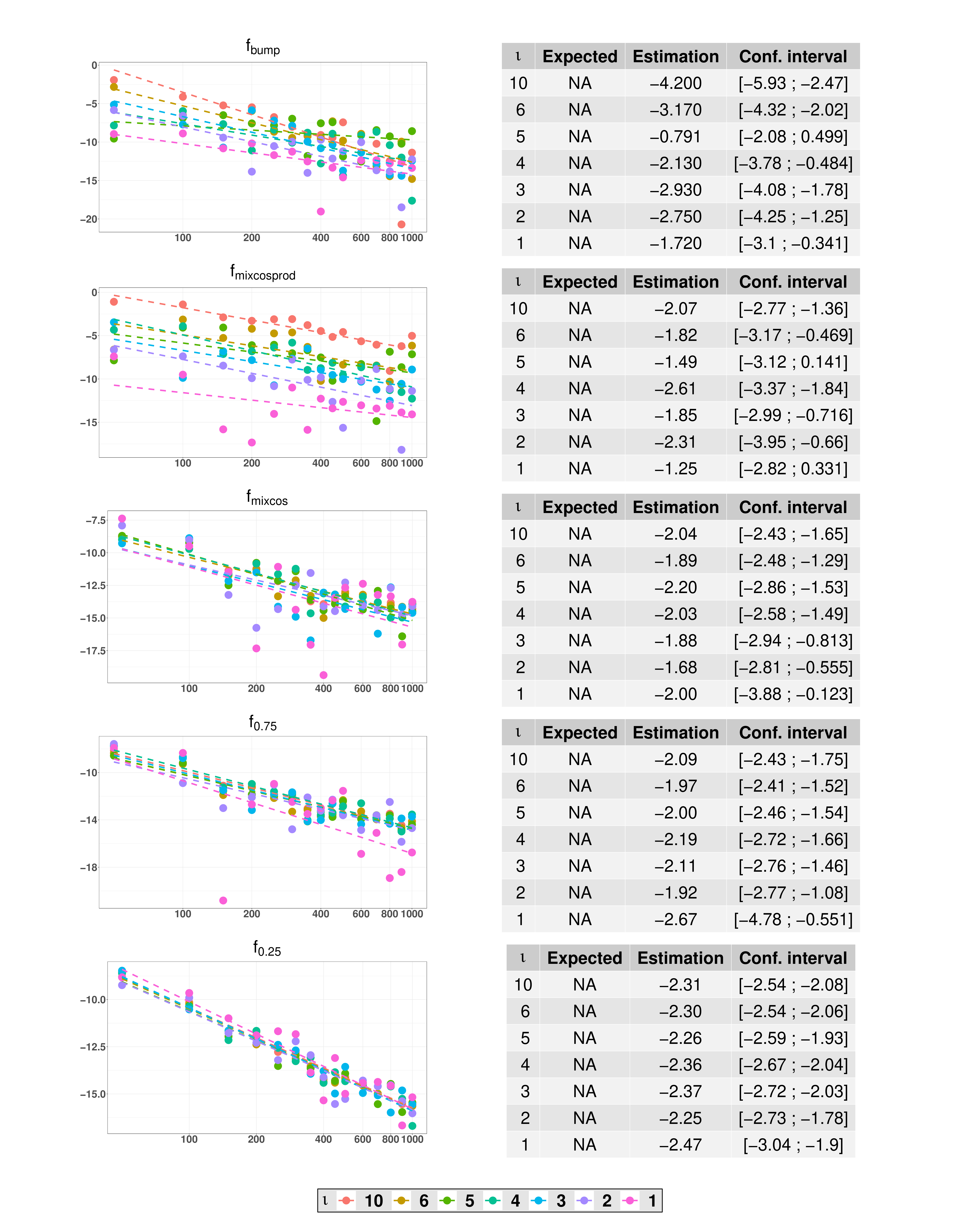}
	\caption{Summary of experiments in which integrals of $\iota$-dimensional functions are estimated by projecting a single 10-dimensional \texttt{Sobol} design ($\iota=1\dots 6,10$).}
	\label{fig:sobol3}
\end{figure}

\begin{figure} 
	\vspace*{-2.5cm}\includegraphics[scale=.16]{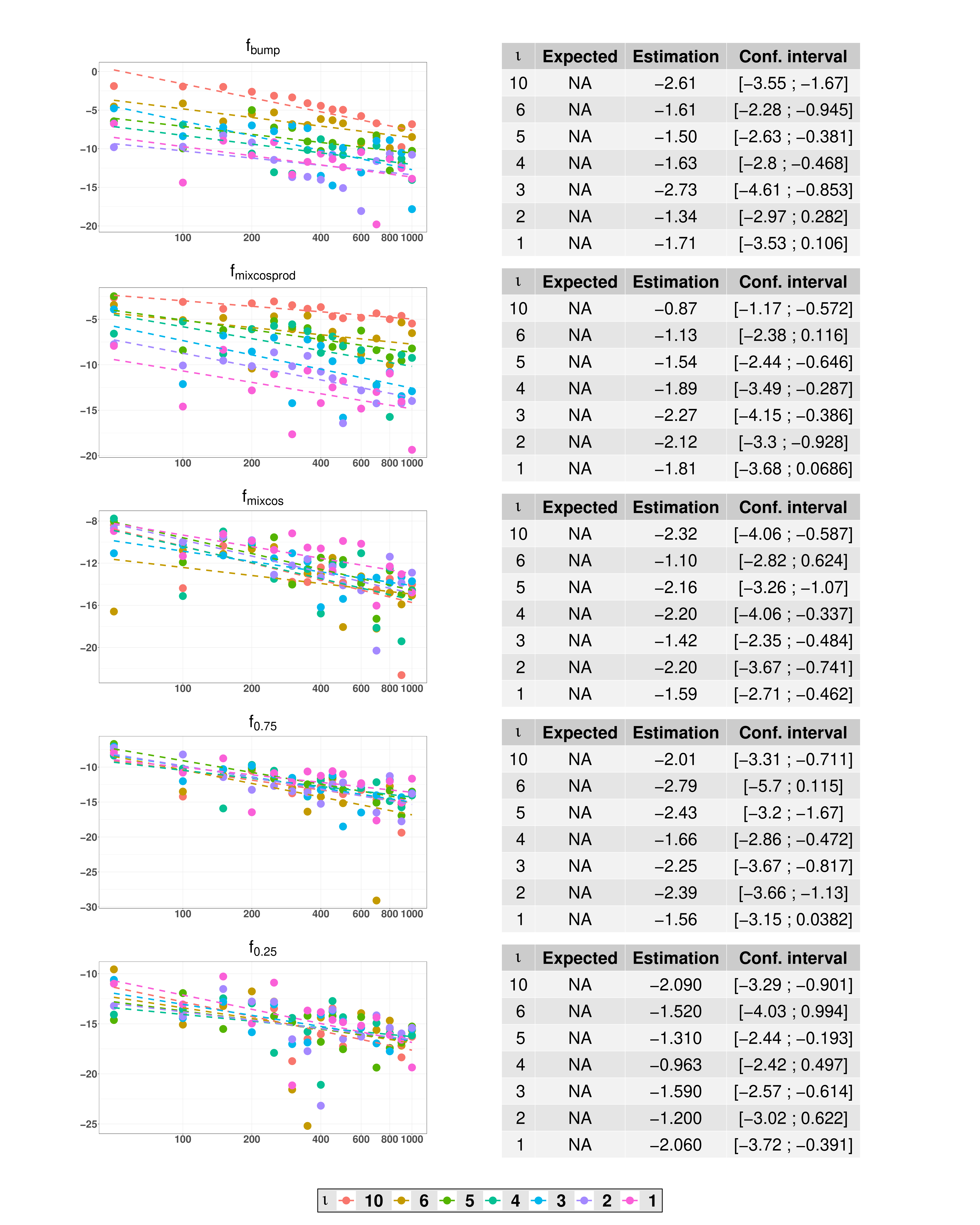}
	\caption{Summary of experiments in which integrals of $\iota$-dimensional functions are estimated by projecting a single 10-dimensional \texttt{Halton} design ($\iota=1\dots 6,10$). }
	\label{fig:halton3}
\end{figure}

{Figures \ref{fig:crude1}-\ref{fig:halton3} display simulation results for the logarithms of the empirical of MSE in terms of $\log(N)$ of estimates of $d$-dimensional integrals ($d=1,2,\dots,6,10$, Figures~\ref{fig:crude1}-\ref{fig:halton1}) or $\iota$-dimensional integrals using point projected point patterns generated in dimension $d=6$ (Figures~\ref{fig:crude2}-\ref{fig:halton2}) or $d=10$ (Figures~\ref{fig:crude3}-\ref{fig:halton3}). Results are presented for the five test functions given by~\eqref{eq:fbump}-\eqref{eq:fgamma} and for the \texttt{crude MC}, \texttt{stratified MC}, \texttt{maximinLHS}, \texttt{Sobol} and \texttt{Halton} designs, see Section~\ref{sec:comparisons} for more details.}

\end{document}